\documentclass[12pt]{article}
\usepackage[margin=0.9in]{geometry}
\usepackage[utf8]{inputenc}
\usepackage{bm}
\usepackage{bbm}
\usepackage{stmaryrd}
\usepackage{amsmath}
\usepackage{amssymb}
\usepackage{amsthm}
\usepackage{xcolor}
\usepackage{graphicx}
\usepackage{ulem}
\usepackage{url}
\usepackage{subcaption}
\usepackage{cite}
\usepackage{authblk}

\usepackage{nicematrix}

\NiceMatrixOptions{cell-space-limits = .2pt}

\newcommand{\e}{{\rm e}}
\newcommand{\tr}{\mathrm{Tr}}
\newcommand{\h}{{\mathcal H}}
\newcommand{\wM}{\widetilde{\mathcal M}}
\newcommand{\s}{{S}}
\renewcommand{\a}{{A}}
\newcommand{\U}{{\mathcal U}}
\newcommand{\W}{{\mathcal W}}
\newcommand{\M}{{\mathcal M}}

\newcommand{\bbbone}{\mathchoice {\rm 1\mskip-4mu l} {\rm 1\mskip-4mu l}
{\rm 1\mskip-4.5mu l} {\rm 1\mskip-5mu l}}

\newtheorem{prop}{Proposition}
\newtheorem{lem}{Lemma}

\newtheorem{theorem}{Theorem}

\begin{document}

\title{Non-Markovianity in collision models with initial intra-environment correlations}

\author[1,2]{Graeme Pleasance\footnote{gpleasance1@gmail.com (corresponding author)}}
\author[3]{\'Angel E. Neira\footnote{aeneira@mun.ca}}
\author[3]{Marco Merkli\footnote{merkli@mun.ca}}
\author[2,4]{Francesco Petruccione\footnote{francesco.petruccione@nithecs.ac.za}}

\affil[1]{\small Department of Physics, University of Stellenbosch, Stellenbosch, 7600, South Africa}
\affil[2]{\small National Institute of Theoretical and Computational Sciences (NITheCS), Stellenbosch, 7600, South Africa}
\affil[3]{\small Department of Mathematics and Statistics, Memorial University of Newfoundland, St. John's, NL, Canada, A1C 5S7}
\affil[4]{\small School of Data Science and Computational Thinking, University of Stellenbosch, Stellenbosch 7600, South Africa}

\maketitle

\begin{abstract}
Collision models (CMs) describe an open system interacting in sequence with elements of an environment, termed ancillas. They have been established as a useful tool for analyzing non-Markovian open quantum dynamics based on the ability to control the environmental memory through simple feedback mechanisms. In this work, we investigate how ancilla-ancilla entanglement can serve as a mechanism for controlling the non-Markovianity of an open system, focusing on an operational approach to generating correlations within the environment. To this end, we first demonstrate that the open dynamics of CMs with sequentially generated correlations between groups of ancillas can be mapped onto a composite CM, where the memory part of the environment is incorporated into an enlarged Markovian system. We then apply this framework to an all-qubit CM, and show that non-Markovian behavior emerges only when the next incoming pair of ancillas are entangled prior to colliding with the system. On the other hand, when system-ancilla collisions precede ancilla-ancilla entanglement, we find the open dynamics to always be Markovian. Our findings highlight how certain qualitative features of inter-ancilla correlations can strongly influence the onset of system non-Markovianity.  
\end{abstract}

\section{Introduction}

Open quantum systems are fundamental to a wide range of quantum technologies---such as quantum sensors and quantum thermal devices \cite{Latune2023,Pleasance2024}---where the behavior of the system of interest inherently depends on its interaction with the environment \cite{Breuer2002}. Under a standard framework in which the environment is assumed to be memoryless and weakly interact with the system, the resulting dynamics may be well approximated by a GKSL master equation describing a Markovian evolution \cite{Gorini1976,Lindblad1976}. Such master equations have the general advantage of being amenable to both analytical and numerical techniques \cite{Johansson2013} as well as offering a physically transparent interpretation of the open dynamics \cite{Plenio1998,Dalibard1992,Gisin1992}. Nonetheless, the growing importance of non-Markovian quantum dynamics \cite{Vega2017} has lead to the development of alternative frameworks for treating open system dynamics with memory, including the hierarchical equations of motion \cite{Tanimura1990,Ishizaki2005}, the pseudomode method \cite{Tamascelli2018, Pleasance2017,Pleasance2021,Pleasance2020,Albarelli2024}, path integral techniques \cite{Weiss2011}, and stochastic approaches \cite{Stenius1996,Piilo2008,Suess2014}.

Among these techniques, collision models (CMs) have emerged as a promising tool for studying open quantum systems in a way that realizes the system-reservoir interaction in a conceptually simple manner \cite{Rau1963, Ciccarello2022}. Rather than assuming the system to interact continuously with the same environment, CMs divide up the environment into a chain of discrete ancillas, which each interact with the system via step-wise collisions. Since this setup offers both a high degree of control and flexibility over the type of evolution it describes \cite{Rybar2012,Filippov2017,Cattaneo2021}, such models have found a broad range of applications spanning quantum optics \cite{Ciccarello2017,Whalen2019,Cuevas2019} and quantum information \cite{Campbell2019,Lorenzo2020}, to quantum thermodynamics \cite{Lorenzo2015,Strasberg2017,Morrone2023,Cusumano2023} and high energy physics \cite{Ribeiro2024}. Along similar lines, CMs are well suited to simulating non-Markovian dynamics due to their capability in realizing memory effects through simple feedback mechanisms \cite{Pellegrini2009,Rybar2012,Ciccarello2013,Ciccarello2013a,McCloskey2014,Kretschmer2016,Filippov2017,Bernardes2014,Bernardes2017,Mascarenhas2017,Lorenzo2017,Campbell2018,Man2018,Whalen2017,Whalen2019,Cuevas2019,Li2021,Morrone2023,Saha2024,Saha2024a}. These mechanisms may be categorized according to various schemes that rely on modifying the basic CM used for Markovian dynamics---either by allowing multiple system-ancilla collisions to occur at each step \cite{Pellegrini2009,Whalen2017,Whalen2019}, or by changing the properties of the environment. Under the latter approach, memory effects may be introduced via environments featuring internal ancilla-ancilla (AA) collisions \cite{Ciccarello2013,Ciccarello2013a,McCloskey2014,Campbell2018,Man2018,Saha2024}, or with initially correlated states \cite{Rybar2012,Bernardes2014,Bernardes2017,Filippov2017,Mascarenhas2017}. Across all such cases, the precise mechanism underpinning the memory is the establishment of correlations between the system and a specific `interacting' portion of the environment. Indeed in Refs. \cite{McCloskey2014,Campbell2018}, the repeated erasure of correlations between the system and the last collided ancilla, which then undergoes AA interactions, was generally shown to alter the non-Markovianity of the open dynamics. On the other hand, ancillas which the system has already collided with, but that have no involvement in later interactions (the `non-interacting' portion of the environment), may be completely neglected from the open dynamics despite possibly retaining correlations with the system. 

Based on this link between system-environment correlations and memory effects, Campbell et.~al.~\cite{Campbell2018} recently introduced the notion of `memory depth' for non-Markovian CMs with AA collisions that enacts a partitioning of the environment into its memory and non-memory parts. The memory depth quantifies the size of the memory in terms of the interacting portion of ancillas that become correlated with the system following consecutive AA collisions---i.e. those whose correlations cannot be neglected without generally changing the open dynamics---and hence must be accounted for in the construction of the system dynamical map. As per \cite{Campbell2018}, this suggests an effective Markovian description of the open dynamics in which the system and memory are treated as a composite system interacting with the remaining non-memory part \cite{Lorenzo2017}. Conceptually, such a description is related to the idea of non-Markovian to Markovian mappings employed more widely in the open quantum systems literature \cite{Tamascelli2018, Pleasance2017,Pleasance2021,Pleasance2020}.

\begin{figure}[t!]
	\centering
        \includegraphics[scale=0.45]{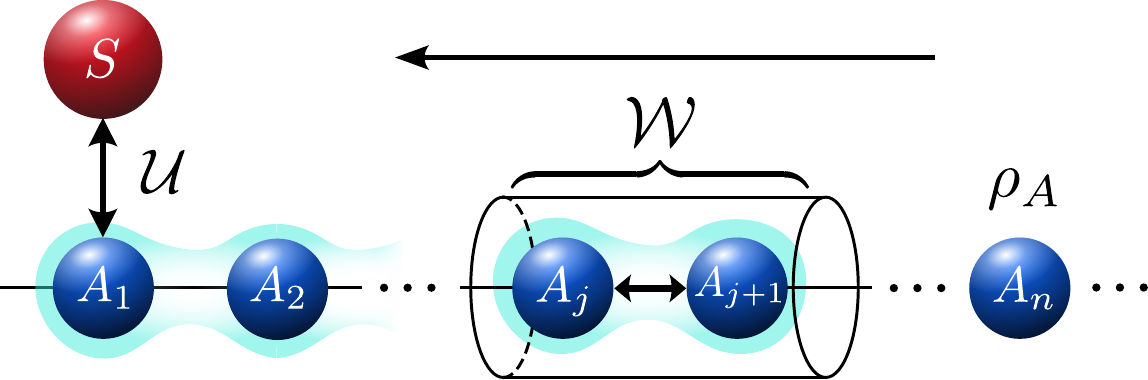}
	\caption{{\bf The model.} An environment $E$ 
    of identical ancillas $A_1,...,A_n$, each in the state $\rho_A$, travels to the left; they reach a mechanism whereby groups of size $L$ become sequentially correlated through an operation $\W$ ($L=2$ in the schematic). After exiting the `correlation tube', overlapping groups of $L$ ancillas end up correlated. The system $S$ then interacts with each ancilla, one by one, via a unitary operation $\U$.}
    \label{fig1}
\end{figure}

In this work, we devise a similar strategy to Ref. \cite{Campbell2018} for treating CMs with correlated environment states, which is used to analyze the role inter-ancilla entanglement plays in the non-Markovianity of an all-qubit CM. To do so, we introduce a class of non-Markovian CMs whereby overlapping groups of consecutive ancillas are correlated in sequence {\it before} colliding with the system; see Fig.~\ref{fig1}. This differs from the scheme of Ref. \cite{Campbell2018} where SA and AA collisions occur in alternating sequence starting with initially uncorrelated ancillas. At the same time, our approach bridges the gap between CMs with correlated environments \cite{Rybar2012,Bernardes2014,Bernardes2017,Filippov2017,Mascarenhas2017} and their experimental implementation \cite{Bernardes2016, Jin2015, Cuevas2019}, in that our scheme accounts for the physical preparation of the initial environment state. We demonstrate that this setup allows for an effective Markovian description of the open dynamics in terms of a {\it composite collision model} (CCM) akin to \cite{Campbell2018,Lorenzo2017}, thereby reflecting an analogous partitioning of the environment into memory and non-memory parts. However, contrary to \cite{Campbell2018}, the construction of the dynamical map is shown to differ not only on the order in which AA and SA collisions are applied, but also the portion of environmental ancillas that the system effectively interacts with. 

For an all-qubit CM with sequentially correlated AA pairs, it is further established that entanglement between ancillas within the interacting portion of the environment is {\it necessary} for the emergence of non-Markovian behavior. Moreover, we show that non-Markovianity appears precisely because such ancilla pairs are entangled before colliding with the system. Indeed, when the ancillas are instead correlated after the system collides with the first ancilla, then the dynamics is entirely Markovian, regardless of the degree of AA entanglement.

The remainder of this paper is organized as follows. In Sec.~\ref{sec2}, we introduce the CM setup displayed in Fig.~\ref{fig1}, where intra-environment correlations are generated via a family of unitary maps applied to consecutive groups of ancillas. In Sec.~\ref{sec3} we demonstrate how the open dynamics of this setup can be mapped onto a composite CM,
which incorporates the memory part of the environment into an enlarged Markovian open system. We then in Sec.~\ref{sec4} apply our result to an all-qubit CM and study the connection between the degree of AA entanglement of relevant ancilla pairs and the non-Markovianity of the open dynamics. Finally, conclusions and outlook are presented in Sec.~\ref{sec5}.

\section{Basic collision model}
\label{sec2}

To fix the notation, we start by outlining the basic setup used to describe an open system $\s$ interacting with a memoryless environment of uncorrelated ancillas \cite{Ciccarello2022}. In such a model the environment $E$ consists of $n$ identical ancillas $\a_j$ ($j=1,\ldots,n$) initialized in the state $\rho_{E}(0)=\rho_{\a_1}\otimes\cdots\otimes \rho_{\a_n}$, where the Hilbert spaces of the system and the $j$-th ancilla are denoted as $\h_\s$ and $\h_{\a_j}$, respectively. Since the ancillas are identical, the index $j$ is simply a label for a copy of the same Hilbert space $\h_\a$. The system and environment interact through pairwise collisions between $\s$ and each $\a_j$, where each collision is described by a unitary operator $U$ acting on $\h_\s\otimes\h_\a$. We shall denote by $U_{\s j}$ the equivalent actions on $\s$ and the ancilla $\a_j$ when several ancillas are present, i.e.~when the total Hilbert space is $\h_\s\otimes\h_{\a_1}\otimes \cdots\otimes\h_{\a_n}$, and 
\begin{equation}
    \rho\mapsto\mathcal{U}_{\s j}\rho=U_{\s j}\rho U^{\dagger}_{\s j}
\end{equation}
the collision map between $\s$ and $\a_j$ acting on density matrices $\rho$ of the system plus $n$ ancillas.

Let us define the single-collision map
\begin{equation}\label{basic_cm}
	\rho_\s \mapsto \Lambda[\rho_\s] = \tr_\a[U(\rho_\s\otimes\rho_\a)U^{\dagger}].
\end{equation} 
It is a completely positive and trace-preserving (CPTP) map on the set of all system density matrices. For an initially factorized state $\rho(0) = \rho_\s(0)\otimes\rho^{\otimes n}_\a$, the reduced system state after $n$ collisions $\rho_\s(n)=\tr_{\a_1...\a_n}[\mathcal{U}_{\s n}\cdots\mathcal{U}_{\s 1}\rho(0)]$ is obtained by iterating $n$ times the map $\Lambda$, 
\begin{equation}
\label{basic_cm_dynamics}
	\rho_\s(n) = \Lambda^n[\rho_\s(0)].
\end{equation}
This implies the system dynamics to be Markovian insofar that the composite map $\Lambda_n\equiv\Lambda^n$ implementing the evolution $\rho_\s(0)\mapsto \rho_\s(n)=\Lambda_n\rho_S(0)$ is divisible, i.e., $\Lambda_n$ can inherently be written as a composition of CPTP maps \cite{Wolf2008,Rivas2014}. At the same time, the memoryless property of Eq.~\eqref{basic_cm_dynamics} may also be understood from fact that it describes the repeated interaction between $\s$ and a {\it single} ancilla $\a$, whose correlations with $\s$ can be erased at each step without affecting the future system evolution (that is, no memory of the system state is retained across successive collisions).

\section{Non-Markovian collision model with correlated ancillas} 
\label{sec3}

As previously noted, the realization of non-Markovian dynamics relies on modifying the basic setup above according to one of several schemes. Two common schemes employed in the literature are: 
\begin{itemize}
\item[(i)] AA collisions: After $\s$ has collided with the ancilla $\a_j$, and before $\s$ collides with the next one $\a_{j+1}$, the two ancillas $\a_j$ and $\a_{j+1}$ interact via a unitary operator $W_{j+1,j}$ \cite{Ciccarello2013, Ciccarello2013a,McCloskey2014,Campbell2018,Man2018,Li2021,Morrone2023}. As such, $A_{j+1}$ already carries memory of the system state before colliding with $S$. In this scheme one can also collide groups of more than two ancillas; the total `length' of each group (number of connections between first and last ancilla) is called the memory depth \cite{Campbell2018}.  

\item[(ii)] Intra-environment correlations: The initial environment state $\rho_E(0)$ is replaced by one that is nonfactorizable, reflecting classical or quantum correlations between ancillas \cite{Rybar2012,Bernardes2014,Filippov2017,Mascarenhas2017}.
\end{itemize}
In the current work, we introduce a different approach to the latter scheme (ii) that closely resembles (i), in which correlations between ancillas are generated operationally. In particular, we consider initial states in which overlapping groups of $L$ ancillas are correlated in sequence before colliding with $S$ (see Fig.~\ref{fig1}). We proceed to formalize the model as follows. First, let $W$ define a unitary acting on any fixed number of $L\ge 1$ ancillas, and let $W_{[j+L-1,j]}$ denote $W$ acting on a string of $L$ ancillas with indices $j,j+1,\ldots,j+L-1$ (while acting as the identity on all other ancillas). We further denote by $\W_{[j+L-1,j]}$ and $\W$ the actions of $W_{[j+L-1,j]}$ and $W$ lifted to density matrices of $\s$ and $n$ ancillas,
\begin{equation}
\W \rho = W \rho W^\dag,\qquad \W_{[L+j-1,j]} \rho = W_{[L+j-1,j]} \rho W_{[L+j-1,j]}^\dag.
\end{equation}
Thus, the initial states we consider are given as
\begin{equation}
\label{m12}
\rho(0) = \rho_\s(0)\otimes \W_{[n+L-1,n]}\cdots\W_{[L+1,2]}\W_{[L,1]}\, \big(\rho_\a^{\otimes n+L-1}\big).
\end{equation} 
By setting $L=1$, it is clear that \eqref{m12} reduces to the same class of product states associated with memoryless CMs. In this regard, we refer to $L$ as the {\it correlation length} of the environment in analogy to the memory depth introduced in \cite{Campbell2018} (see further discussion below). We remark that it is also feasible to allow for more general types of operations $\mathcal W$ represented by non-unitary CPTP maps. 

Starting from the initial condition \eqref{m12}, the full system-reservoir state evolves as
\begin{align}
\rho(n) &= \mathcal{U}_{\s n}\cdots \mathcal{U}_{\s 1}\big(\rho_\s(0)\otimes \mathcal{W}_{[n+L-1,n]}\cdots\mathcal{W}_{[L,1]}\,\rho^{\otimes n+L-1}_\a\big) \nonumber\\
	   &= \mathcal{U}_{\s n}\mathcal{W}_{[n+L-1,n]}\cdots\mathcal{U}_{\s 1}\mathcal{W}_{[L,1]}\big(\rho_\s(0)\otimes \rho^{\otimes n+L-1}_\a\big),
\end{align}
such that the reduced system state after $n$ collisions reads
\begin{align}
\label{reduced_state}
\rho_\s(n) &= \tr_{\a_1...\a_{n+L-1}}\big[\mathcal{U}_{\s n}\mathcal{W}_{[n+L-1,n]}\cdots\mathcal{U}_{S1}\mathcal{W}_{[L,1]}\big(\rho_\s(0)\otimes \rho^{\otimes n+L-1}_\a\big)\big]\nonumber\\
&\equiv\Lambda_n[\rho_\s(0)].
\end{align}
In contrast to \eqref{basic_cm_dynamics}, the map $\Lambda_n$ is generally non-Markovian since it is not guaranteed to be factorizable into products of CPTP maps. 

\smallskip

{\it Remark.} 
The operators $U$ and $W$ commute exactly when $\mathcal{U}$ and $\mathcal{W}$ do. In this case, we have 
\begin{equation}
\U_{\s n}\cdots \U_{\s 1} \W_{[n+L-1,n]}\cdots\W_{[L,1]}
=\W_{[n+L-1,n]}\cdots\W_{[L,1]} \U_{\s n}\cdots \U_{\s 1}, \quad [U,W]=0.
\label{m14.1}
\end{equation}
When tracing over the ancilla degrees of freedom in \eqref{reduced_state}, the action of each $\mathcal{W}$ now disappears since these superoperators represent a conjugation with a unitary. It then follows that the reduced system dynamics is independent of the intra-environment correlations provided these are generated by an operator $W$ that commutes with the $\s-\a$ interaction.

\begin{figure}[t!]
	\centering
        \includegraphics[scale=0.35]{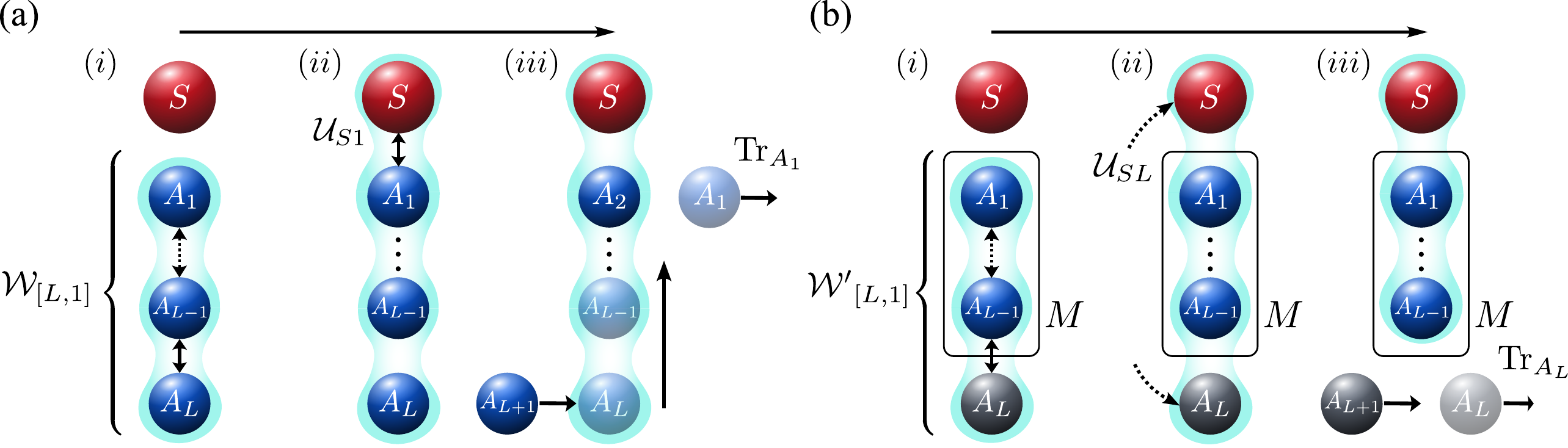}
	\caption{{\bf The map $\mathcal M$ (acting on $\rho_S\otimes\rho^{\otimes L-1}_A$) viewed in two different ways according to \eqref{20m} and \eqref{anfin}, with the chronological order of operations labeled by {\it (i)-(iii)}.} {\bf (a)} The map represented by \eqref{20m}. In the first step {\it (i)}, the ancillas $\a_1,...,\a_L$ are correlated via $\W_{[L,1]}$ (the case where $\mathcal W$ performs pairwise collisions is shown). This is followed by {\it (ii)} $\s$ colliding with $\a_1$, and {\it (iii)} $\a_1$ being traced out. The ancillas are then shifted: $\a_2$ takes the role of the previous $\a_1$ and so on, and a new ancilla $\a_{L+1}$ is introduced in place of $\a_L$. {\bf (b)} The map represented by \eqref{anfin}. A different sequence of operations {\it (i)-(iii)} is now applied, with $\W_{[L,1]}$ and $\U_{S1}$ replaced by $\W'_{[L,1]}$ and $\U_{SL}$, and where the partial trace is taken over the incoming ancilla $A_L$. As such, $S$ interacts with the same group of ancillas $M$ with repeated applications of \eqref{anfin}.}
    \label{fig2}
\end{figure}

\subsection{Mapping onto Markovian composite collision model}
\label{sec3.1}

Following the ideas of Refs. \cite{Campbell2018,Lorenzo2017}, we now demonstrate how the system dynamics $\Lambda_n$ can be realized as a Markovian process of an enlarged system containing $S$ and $L-1$ ancillas, an approach known as Markovian embedding. The main advantage of such an embedding is that the dynamics of the enlarged system may be treated using techniques applicable to maps with the same composite structure as \eqref{basic_cm_dynamics}. On the other hand, the dimension of this system grows exponentially with the correlation length $L$; for example, if the ancillas are of the dimension ${\rm dim}(\h_\a)=d$, then the dimension of the enlarged system is ${\rm dim} (\h_\s\otimes\h_\a^{\otimes L-1}) =  {\rm dim}(\h_\s)\cdot d^{L-1}$. Hence, such a mapping in practice is limited to small values of $L$.

Our first result is that the dynamical map $\Lambda_n$, \eqref{reduced_state}, may be expressed as the $n$-th iteration of another map $\M$ that transforms between states of $S$ and {\it different} groups of $L-1$ consecutive ancillas, as depicted schematically  in Fig.~\ref{fig2}(a) (below the index of the ancillas is suppressed since the state of each ancilla is represented on a duplicated Hilbert space $\h_A$). To state this formally, we introduce $\mathcal S(\h_S\otimes\h^{\otimes L-1}_A)$ as the space of all density matrices of $S$ plus $L-1$ ancillas, and denote by $\vartheta$ an arbitrary element of $\mathcal S(\h_S\otimes\h^{\otimes L-1}_A)$. We then have the following result.
\begin{theorem}
\label{thm1}
Let $\M : \mathcal{S}(\h_S\otimes\h^{\otimes L-1}_A)\rightarrow \mathcal{S}(\h_S\otimes\h^{\otimes L-1}_A)$ be the following CPTP map,
\begin{equation}
\label{20m}
\vartheta\mapsto\M[\vartheta] = \tr_{\a_1}\Big( \U_{\s 1}\W_{[L,1]} (\vartheta\otimes\rho_\a)\Big).
\end{equation}
Then the reduced system state \eqref{reduced_state} after $n$ collisions is expressed as
\begin{equation}
\label{m21.1}
\rho_\s(n) = \tr_{\a_{L-1}\ldots\a_1} \M^n[\rho_\s(0)\otimes\rho^{\otimes L-1}_\a],
\end{equation}
where $\M^n$ is the $n$-th power of $\M$, and the trace in \eqref{m21.1} is taken over all $L-1$ ancillas.
\end{theorem}

We present a proof of Theorem \ref{thm1} in Sec. \ref{deriv:21.1}. To clarify the interpretation of \eqref{m21.1}, consider the $L=2$ case where AA pairs are sequentially correlated. In this scenario, the map \eqref{20m} describes a sequence of three operations
involving the system $\s$ and two ancillas, say $\a_1$ and $\a_2$. These ancillas are first correlated through the operation $\mathcal{W}_{[2,1]}$. This is followed by a collision between $\s$ and $\a_1$, and finally the ancilla $\a_1$ is traced out. The resulting $\M[\vartheta]$ now represents the state of $\s$ plus ancilla $\a_2$. Applying the same map again then takes the state of the enlarged system $\s+\a_2$, and implements the same three operations to obtain the state of $\s+\a_3$. Continuing this process $n$ times we obtain $\M^n[\vartheta]$, i.e. the state of $\s$ and ancilla $\a_{n+1}$, so that tracing over $A_{n+1}$ recovers the reduced system state from \eqref{reduced_state}. We also highlight that the map $\mathcal M$, \eqref{20m} itself depends on the ancilla state $\rho_\a$, which is considered arbitrary but fixed.
\medskip

{\bf Composite system representation of the open dynamics.} We next prove that the description of the dynamics given by \eqref{m21.1} coincides with an alternative representation whereby correlations are generated between $A_{j+L-1}$ and the {\it same} group of $L-1$ ancillas $M = \{A_1,...,A_{L-1}\}$, rather than between a different subset of ancillas at each iteration. To this end, let $\rho\mapsto \mathcal{S}_{k,l}\rho = S_{k,l}\rho S_{k,l}$ define the swap map that interchanges the states of the  ancillas with indices $k$ and $l$, according to (omitting the tensor product symbols)  
\begin{equation}
\mathcal S_{k,l}[\rho_{\a_1}\cdots\rho_{\a_k} \cdots\rho_{\a_l}\cdots \rho_{\a_n}] = \rho_{\a_1}\cdots\rho_{\a_l} \cdots\rho_{\a_k}\cdots \rho_{\a_n}.
\end{equation}
We then define the operator
\begin{equation}
    \W_{[L,1]}'=\mathcal{S}_{L,L-1}\cdots\mathcal S_{2,1}\W_{[L,1]}
\end{equation} 
and show in Sec.~\ref{sec:DerivAnfin} the following result.

\begin{prop}
\label{prop1-1}
The map $\M$ defined in \eqref{20m} may alternatively be written in the form 
\begin{equation}
\label{anfin}
\M[\vartheta]=\tr_{\a_L}\Big(\U_{\s L}\W_{[L,1]}' (\vartheta\otimes\rho_\a)\Big),
\end{equation}
where the trace is taken over the ancilla $A_L$.
\end{prop}

\begin{figure}[t!]
    \centering
    \includegraphics[scale=0.35]{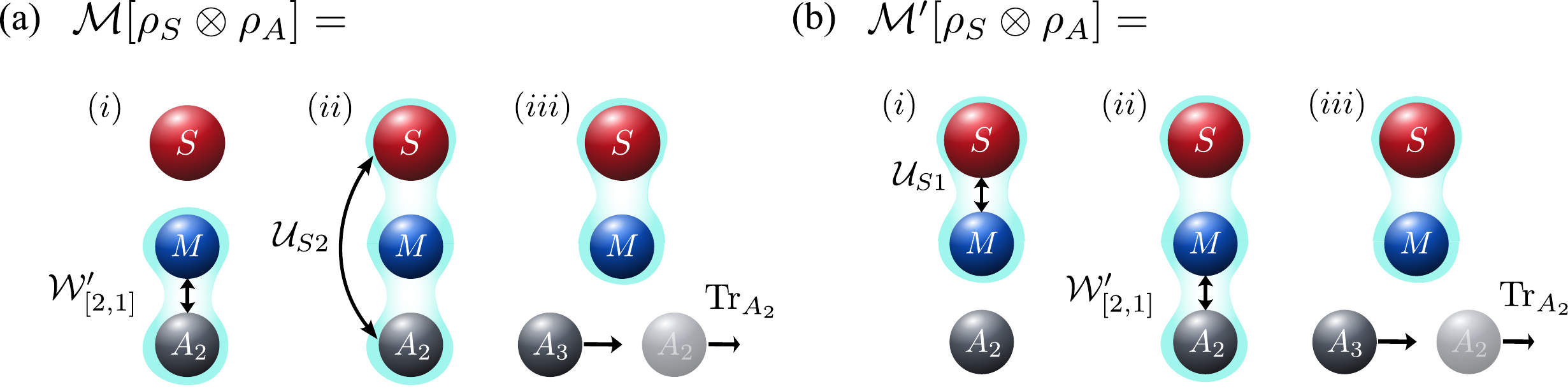}
    \caption{{\bf The maps $\mathcal M$ and $\mathcal M'$ schematically represented in the case of $L=2$.} {\bf (a)} The system $S$ interacts with the memory ancilla $M$ ($A_1$) and non-memory ancilla $A_2$ as per Fig.~\ref{fig2}(b). {\bf (b)} The map $\mathcal M'$ introduced in \cite{Ciccarello2013, Ciccarello2013a,McCloskey2014,Campbell2018} to describe the dynamics of the analogous CCM in which there are no initial correlations, but where AA collisions occur via $\mathcal{W}'_{[2,1]}$ {\it after} $S$ collides with $M$. In contrast to $\M$, the order of the operations {\it (i)} and {\it (ii)} is reversed, in addition to $S$ colliding directly with the memory ancilla.}
    \label{fig3}
\end{figure}

The interpretation of the system dynamics as viewed by \eqref{anfin} is that an incoming ancilla $A_{L}$ is initially correlated with a collection $M$ of $L-1$ ancillas via $\mathcal{W}'_{[L,1]}$, which then collides with $S$, before being traced out and replaced by another ancilla $A_{L+1}$; see Fig.~\ref{fig2}(b). Notably, because it is the same incoming ancilla that is traced out after each SA collision, the ancillas $M$ can be interpreted as the {\it memory part} of the environment whose correlations with the system generally play a nontrivial role in the system evolution \cite{Campbell2018}. On the other hand, the ancillas not contained in this group represent the non-memory part insofar that their correlations with $S$ can be erased at each step without affecting the system dynamics. We recall that a similar interpretation was applied to the map $\Lambda$ defined in the basic CM, \eqref{basic_cm_dynamics}. Our result---that the generally non-Markovian system dynamics $\Lambda_n$ can be obtained as the reduction of the Markovian dynamics $\mathcal{M}$ of a larger (system-ancilla) complex---establishes what is referred to in the literature as a Markovian embedding, or equivalently a non-Markovian to Markovian mapping of the open dynamics \cite{Tamascelli2018,Pleasance2020,Pleasance2021}. We emphasize that such a mapping holds for arbitrary values of the parameter $L$, in addition to being valid for arbitrary operations $\mathcal U$ and $\mathcal W$ represented by CPTP maps.

Naturally, since the correlation length quantifies the size of the memory part retaining essential correlations with the system, it adopts the same role as the memory depth from Ref. \cite{Campbell2018} for CMs whose non-Markovianity stems from AA collisions. This in turn establishes a direct correspondence between the two mechanisms (i) and (ii), in which the correlations forming the memory are either dynamically generated through AA collisions \cite{McCloskey2014,Campbell2018}, or through intra-environment correlations present in the initial state (with a specific correlation structure determined by \eqref{m12}). It should be noted, however, that the role of the memory in these two schemes differs based on the structure of the CCMs governing the respective system dynamics. In particular, the map $\mathcal M'$ describing the evolution of the composite system $S+A_1$ for a memory depth $L=2$ is given as\footnote{To maintain consistency with our definition of $L$, the map \eqref{M'} corresponds to a memory depth of $L=2$, whereas in Ref. \cite{Campbell2018} it corresponds to a memory depth of one. In both definitions the memory is represented by the same ancilla $A_1$.} \cite{Campbell2018}
\begin{equation}\label{M'}
    \vartheta\mapsto \M'[\vartheta] = \tr_{A_2}\Big(\mathcal{W}'_{[2,1]}\mathcal{U}_{\s 1}(\vartheta\otimes\rho_\a) \Big).
\end{equation}
According to this map, the ancilla $A_1$ first collides with $S$, which then collides with the non-memory part of the environment, as shown in Fig.~\ref{fig3}(b). On the other hand, for our setup with $L=2$, shown in Fig.~\ref{fig3}(a), the order of these operations is {\it reversed}, in addition to the fact that the system only collides with the memory $A_1$ through the non-memory part of the environment. We demonstrate below how these differences in the ordering of SA and AA collisions can significantly impact the non-Markovianity of the dynamics.

\section{All-qubit collision model}
\label{sec4}

In this section, we present our main results on the role of AA entanglement on the non-Markovianity of an open two-level system. For simplicity, we focus on the case where the ancillas $A_j$ are also two-level systems, i.e. $\h_\s=\h_{\a_j}=\mathbb{C}^2$. We further assume a correlation length of $L=2$ and take the SA and AA collision operators to be
\begin{equation}
\label{13.g}
	U_{\s j} = e^{-i\tau \sigma_x\otimes\sigma^{(j)}_x }, \quad W_{[j+1,j]} = e^{-i\varepsilon \sigma^{(j+1)}_z\otimes\sigma^{(j)}_z}.
\end{equation}
Here, $\sigma_{x,y,z}$ ($\sigma^{(j)}_{x,y,z}$) are the Pauli spin-$\frac{1}{2}$ operators of the system ($j$-th ancilla), and $\tau,\varepsilon\ge0$ define the interaction phase and entanglement phase, respectively. 

\subsection{Reduced system dynamics and decoherence function}

The explicit form of the map $\mathcal M$, \eqref{20m}, is given by
\begin{equation}
\label{m46'}
\M[\vartheta] =\tr_{\a_1}\Big(e^{-i\tau \sigma_x\otimes\sigma^{(1)}_x\otimes\bbbone} e^{-i\varepsilon\bbbone\otimes\sigma^{(2)}_z\otimes\sigma^{(1)}_z} (\vartheta\otimes\rho_\a)e^{i\varepsilon\bbbone\otimes\sigma^{(2)}_z\otimes\sigma^{(1)}_z}e^{i\tau \sigma_x\otimes\sigma^{(1)}_x\otimes\bbbone} \Big),
\end{equation}
where $\vartheta$ is the state of $\s$ and a given ancilla $\a_1$. Hence, by defining $\vartheta_n=\M^n[\rho_\s(0)\otimes\rho_\a]$, the reduced system state \eqref{m21.1} takes the form 
\begin{equation}
\label{m21.1'}
\rho_\s(n) = \tr_{\a_1,\a_2} \Big(e^{-i\tau \sigma_x\otimes\sigma^{(1)}_x\otimes\bbbone} e^{-i\varepsilon\bbbone\otimes\sigma^{(2)}_z\otimes\sigma^{(1)}_z} (\vartheta_{n-1}\otimes\rho_\a)e^{i\varepsilon\bbbone\otimes\sigma^{(2)}_z\otimes\sigma^{(1)}_z}e^{i\tau \sigma_x\otimes\sigma^{(1)}_x\otimes\bbbone} \Big).
\end{equation}
Let $|\pm\rangle$ ($|\pm\rangle_A$) denote the eigenstates of $\sigma_x$ ($\sigma^{(j)}_x$). Using \eqref{m21.1'}, it is straightforward to verify that the populations $\langle \pm|\rho_S(n)|\pm\rangle$ are invariant in the basis $\{|+\rangle, |-\rangle\}$, i.e. $\langle \pm|\rho_S(n)|\pm\rangle = \langle \pm|\rho_S(n-1)|\pm\rangle$, and for $\langle +|\rho_S(n)|-\rangle$ to be proportional to $\langle+|\rho_S(n-1)|-\rangle$. Hence, we may write 
\begin{equation}
\label{dephase_map}
\rho_\s(n) = 
	\begin{pmatrix}
		p & D(n)z \\
		D(n)^*z^* & 1-p
	\end{pmatrix}, \qquad (\sigma_x{\rm-basis})
\end{equation}
where $D(n)\in\mathbb{C}$ is the decoherence function of $S$, $D(0)=1$, and $p=\langle +|\rho_\s(0)|+\rangle$, $z=\langle+|\rho_\s(0)|-\rangle$.

By employing a Liouville space representation of the map \eqref{m46'}, the decoherence function $D(n)$ can be computed as a sum of matrix elements of $\M^n$, which in general depends on all parameters $n$, $\varepsilon$, $\tau$, and the initial state $\rho_A$. To show this, we parametrize $\rho_\a$ by its matrix elements in the $\sigma_z$ basis as 
\begin{equation}
\label{m60''}
\rho_\a =
\frac{1}{2}
\begin{pmatrix}
1 +\rho_3 & \rho_1+\rho_2\\
\rho_1-\rho_2 & 1 -\rho_3
\end{pmatrix},\qquad (\sigma_z{\rm-basis}),
\end{equation}
where $\rho_1^2+|\rho_2|^2 +\rho^2_3\le 1$. We then have the following result, proved in Sec. \ref{proofprop2}.

\begin{prop}
\label{prop2}
The decoherence function $D(n)$ in \eqref{dephase_map} is given by 
\begin{equation}
\label{Decoresult}
D(n)=\sum_{m=0}^3\big[(\cos(2\tau)\boldsymbol{A}_1-i\sin(2\tau)\boldsymbol{A}_2)^n\big]_{0m}\, \rho_m,
\end{equation}
where the $\rho_m$, $m=0,\ldots,3$ are given by \eqref{m60''} and  $[\cdots]_{kl}$ denotes the $k,l$ matrix element of the matrices
\begin{equation}
\boldsymbol{A}_1 =
\begin{pmatrix}
1 & 0 & 0 & 0\\
\rho_1\cos(2\varepsilon) & 0 & 0 &-i\rho_2\sin(2\varepsilon)\\
\rho_2\cos(2\varepsilon) & 0 & 0 & -i\rho_1\sin(2\varepsilon)\\
\rho_3 & 0 & 0 & 0
\end{pmatrix},\qquad 
\boldsymbol{A}_2 =
\begin{pmatrix}
0 & \cos(2\varepsilon) & -i\rho_3\sin(2\varepsilon) & 0\\
0 & \rho_1 & 0 & 0\\
0 & \rho_2 & 0 & 0\\
0 & \rho_3\cos(2\varepsilon) & -i\sin(2\varepsilon) & 0 
\end{pmatrix}.
\label{A1A2}
\end{equation}
\end{prop}

For simplicity, we will assume hereon for each ancilla to be intialized in the state 
\begin{equation}
\label{rhoa}
\rho_A=|+\rangle\langle +|_A,
\end{equation}
corresponding to $\rho_0=\rho_1=1$, and $\rho_2=\rho_3=0$. 

\subsection{Effect of AA entanglement on the system non-Markovianity}

{\bf Measure of non-Markovianity.} We now introduce a suitable quantifier of non-Markovianity based on the trace distance measure proposed in \cite{Breuer2009}. The trace distance quantifies the distinguishability of any pair of density matrices $\rho^1$ and $\rho^2$, and is defined as
\begin{equation}
	\mathcal{D}(\rho^1,\rho^2) = \frac{1}{2}\|\rho^1 - \rho^2\|_1,
\end{equation} 
where $\|\cdot\|_1$ is the trace norm $\|X\|_1=\tr\sqrt{X^{\dagger}X}$. The states $\rho^1$ and $\rho^2$ are maximally distinguishable when their trace distance equals one, which occurs if and only if $\rho^1$ and $\rho^2$ are orthogonal. In a Markovian process, the distinguishability between any pair of initial system states $\rho^{1,2}_S(0)$ only ever decreases with increasing collision number $n$. This specifically amounts to a continual loss of information from the system to the environment. On the other hand, non-Markovian effects are identified when information flows from the environment back to the system, corresponding to a temporary increase in the trace distance between successive collisions. 

For the discrete evolution $n\mapsto\rho_S(n)$, this behavior is captured by the the Breuer-Piilo-Laine (BLP) measure \cite{Breuer2009,Laine2010}
\begin{equation}
\label{g22}
	\mathcal{N}_{BLP} = \max_{\rho^1_S(0),\rho^2_S(0)}\sum_{n\in S_+}\Big[\mathcal{D}\big(\rho^1_S(n+1),\rho^2_S(n+1)\big) - \mathcal{D}\big(\rho^1_S(n),\rho^2_S(n)\big)\Big],
\end{equation}
which by definition is maximized over every choice of initial state pair $\{\rho^{1,2}_S(0)\}$. The sum is taken over the set 
\begin{align}
	S_+ &= \big\{n\in\mathbb{N}\,:\,\mathcal{D}\big(\rho^1_S(n+1),\rho^2_S(n+1)\big) - \mathcal{D}\big(\rho^1_S(n),\rho^2_S(n)\big)>0\big\}.
\end{align}
As such, \eqref{g22} is non-zero if and only if there is some initial state pair for which the trace distance temporarily increases. For the all-qubit CM considered above, it is possible to evaluate the trace distance as
\begin{equation}
\label{24.1}
	\mathcal{D}(\rho^1_S(n),\rho^2_S(n)) = \sqrt{(p_1-p_2)^2 + |D(n)|^2|z_1-z_2|^2},
\end{equation}
implying the set $S_+$ to be independent of the initial states $\rho^{1,2}_S(0)$, i.e. 
\begin{equation}
	n\in S_+\,\,\Longleftrightarrow\,\, |D(n+1)|>|D(n)|. 
\end{equation}
Hence, both a necessary and sufficient condition for non-Markovianity in this case is for $|D(n)|$ to increase when going from collision $n$ to $n+1$. 

For qubits, the maximal value in \eqref{g22} is achieved for pairs of states that are antipodal points on the Bloch sphere \cite{Wissmann2012}. The trace distance between any pair of antipodal states can be parametrized in terms of a single parameter $\eta\in[0,1]$,
\begin{equation}
\label{27}
\mathcal{D}\big(\rho^1_S(n),\rho^2_S(n)\big) =\sqrt{\eta +|D(n)|^2(1-\eta)}, \qquad \rho^1_S\perp\rho^2_S.
\end{equation}
The maximization in \eqref{g22} is then a maximization over $\eta$:
\begin{equation}	
\label{27'}
\mathcal{N}_{BLP} = \max_{\eta\in[0,1]}\sum_{n\in S_+}\left(\sqrt{\eta+|D(n+1)|^2(1-\eta)}-\sqrt{\eta+|D(n)|^2(1-\eta)}\right).
\end{equation}

\begin{figure}[h!]
    \centering
    \includegraphics[scale=0.9]{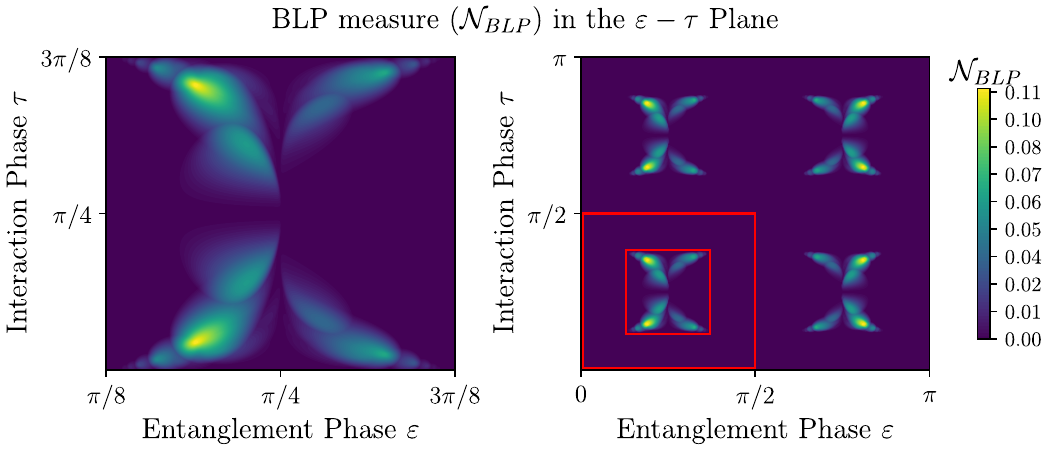}
\caption{{\bf Non-Markovianity measure $\mathcal N_{BLP}$, \eqref{27'}}, evaluated to $n=100$ collisions, for which the non-Markovianity reaches its asymptotic value (increasing $n$ further does not change the results). The ancilla state is $\rho_\a=|+\rangle\langle +|_A$ (see \eqref{rhoa}). The bright regions correspond to a higher degree of non-Markovianity, while the dark regions correspond to parameters $(\varepsilon, \tau)$ for which the behavior is fully Markovian. The right panel shows that $\mathcal{N}_{BLP}$ is symmetric under reflections about $\varepsilon=\frac{\pi}{2}$ and $\tau=\frac{\pi}{2}$. The left panel is an enlargement of the region enclosed by the inner red square in the right panel. } 
\label{fig4}
\end{figure}

\begin{figure}[t!]
	\centering
	\includegraphics[scale=0.65]{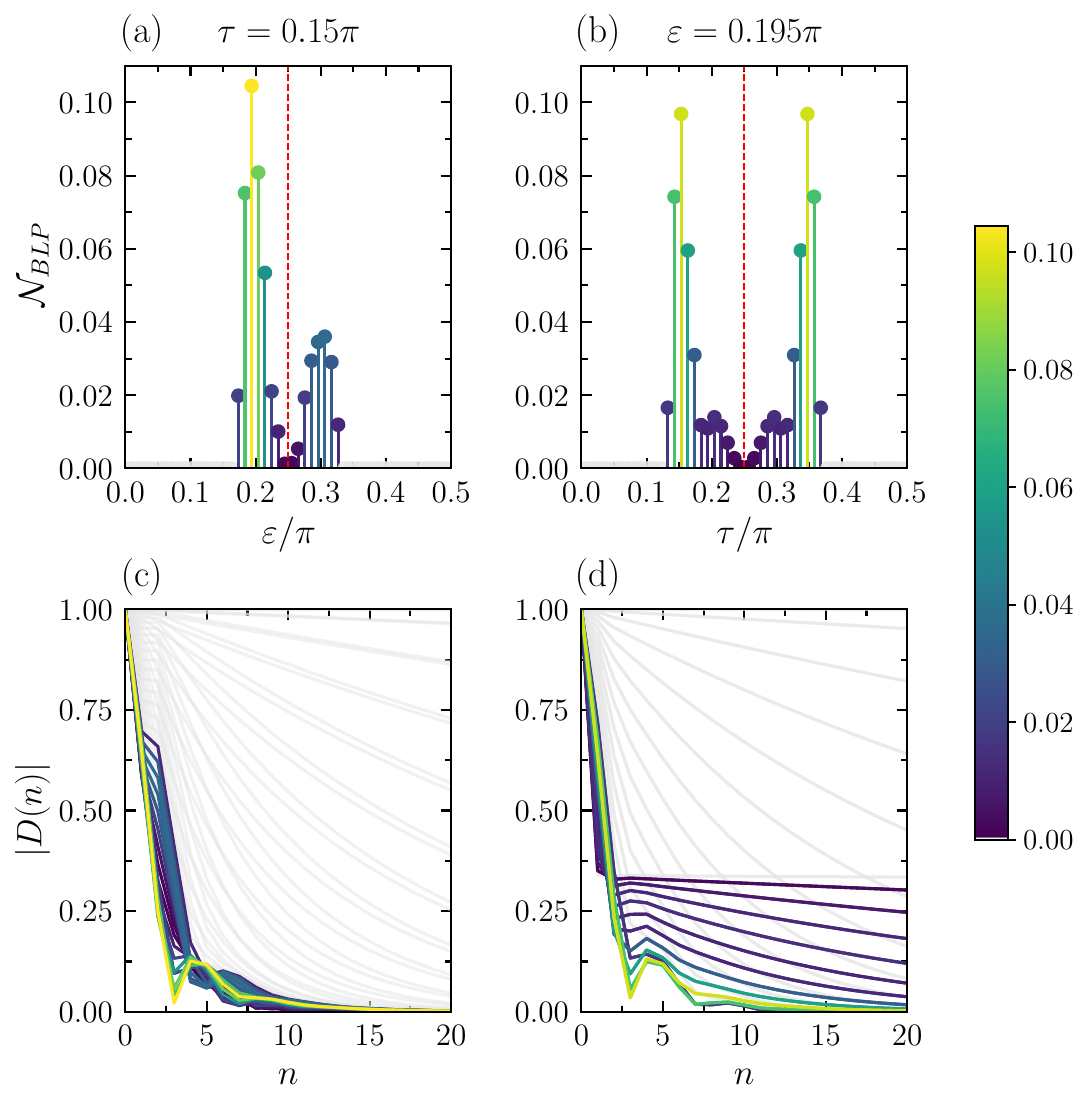}
    \caption{{\bf Snapshots of the non-Markovianity measure} $\mathcal{N}_{BLP}$ shown in Fig. \ref{fig4} for {\bf (a)} $\tau=0.15\pi$, and {\bf (b)} $\varepsilon=0.195\pi$. In both panels, the red dashed lines indicate the values $\varepsilon=\frac{\pi}{4}$ and $\tau=\frac{\pi}{4}$ where the measure vanishes (see \eqref{31'} and \eqref{32'}). {\bf (c), (d) Corresponding dynamics of} $|D(n)|$ {\bf as a function of the collision number} $n$. The gray lines represent instances where $|D(n)|$ decreases monotonically, indicating Markovian behavior.}
    \label{fig5}
\end{figure}

{\bf Numerical results.} In Fig.~\ref{fig4}, we display the measure $\mathcal N_{BLP}$ as a function of $(\varepsilon,\tau)$ for a total of $n=100$ collisions (capping $S_+$ at $n=100$). The lighter areas indicate parts of the parameter space where memory effects play a non-trivial role in the reduced system dynamics, while darker areas represent instances when the dynamics is Markovian (or approximately so). The decoherence function $D(n)$ has symmetries which are inherited by $\mathcal N_{BLP}$ and it suffices to consider the parameter range $(\varepsilon,\tau)=[\frac{\pi}{8}, \frac{3\pi}{8}]\times [\frac{\pi}{8}, \frac{3\pi}{8}]$ depicted in the left panel of Fig.~\ref{fig4}, since the behaviour outside this region can be reconstructed by reflections. Furthermore, the non-Markovianity vanishes within the region enclosed between the red lines in the right panel of the same figure; this occurs trivially on the boundaries where $\varepsilon$ and $\tau$ are either 0 or $\frac{\pi}{2}$. In particular, for $\varepsilon=0$ and $\varepsilon=\frac{\pi}{2},$ no AA correlations are generated and the CM reduces to the basic setup where the system map $\Lambda_n$ has a semigroup structure (see \eqref{basic_cm_dynamics}). It is found that the non-Markovianity is approximately maximal for
\begin{equation}
\label{maxparameters}
\varepsilon_{\rm max}\approx0.195\pi,\qquad \tau_{\rm max}\approx 0.15\pi, \ 0.35\pi,
\end{equation}
for which there are two global maxima (in the considered parameter range) with respect to $\tau$. Additional snapshots of the non-Markovianity measure are presented in Fig.~\ref{fig5} for fixed values of the entanglement and interaction phases. Here, the corresponding dynamics of $|D(n)|$ is also shown (see \eqref{27}).

Comparing now the AA entanglement to the degree of non-Markovianity of the dynamics, we choose to quantify the entanglement between any consecutive pair of ancillas $A_n-A_{n+1}$ immediately {\it before} $A_n$ collides with $S$ via the concurrence \cite{Wootters1998}, 
\begin{equation}\label{concurrence}
    \mathcal{C}(\rho_{A_nA_{n+1}}) =  {\rm max}\{0, \lambda_1(n)-\lambda_2(n)-\lambda_3(n)-\lambda_4(n)\}.
\end{equation}
Here, $\lambda_1(n)\geq\lambda_2(n)\geq\lambda_3(n)\geq\lambda_4(n)$ are the square roots of the eigenvalues of the matrix $\rho_{A_nA_{n+1}}(\sigma^{(n)}_y\otimes\sigma^{(n+1)}_y)\rho^*_{A_nA_{n+1}}(\sigma^{(n)}_y\otimes\sigma^{(n+1)}_y)$, and $\rho^*_{A_nA_{n+1}}$ is the complex conjugate of $\rho_{A_nA_{n+1}}$ (when expressed as a matrix in the $\sigma_z\otimes\sigma_z$ basis). The density matrix $\rho_{A_nA_{n+1}}$ representing the interacting portion of the environment (i.e. the subset of ancillas that are correlated with $S$ before their interaction) before the $n$-th collision may be obtained as 
\begin{align}
    \rho_{A_nA_{n+1}} 
    &= \tr_{SA_1...A_{n-1}}[\U_{Sn-1}\cdots \U_{S1}\W_{[n+1,n]}\cdots\W_{[2,1]}(\rho_S(0)\otimes\rho^{\otimes n+1}_A)] \nonumber\\
    &= \tr_{A_1...A_{n-1}}[\W_{[n+1,n]}\cdots\W_{[2,1]}(\rho_A\otimes\cdots\otimes\rho_A)] \nonumber\\
    &= \W_{[n+1,n]}\Big(\tr_{A_{n-1}}[\W_{[n,n-1]}\cdots\tr_{A_1}[\W_{[2,1]}(\rho_A\otimes\rho_A)]\cdots\otimes\rho_A]\otimes\rho_A\Big). \label{rho_AA}
\end{align}
Note that the second equality follows from the cyclicity of the trace. Hence, since the state $\rho_{A_nA_{n+1}}$ is independent of the $S-A$ interaction, the concurrence \eqref{concurrence} depends only on the initial preparation of the environment through $\W$. In fact, \eqref{rho_AA} is obtained equivalently by taking the trace over all ancillas except $A_n$ and $A_{n+1}$ in the initial state \eqref{m12}. 

\begin{figure}[t!]
    \centering
    \includegraphics[scale=0.65]{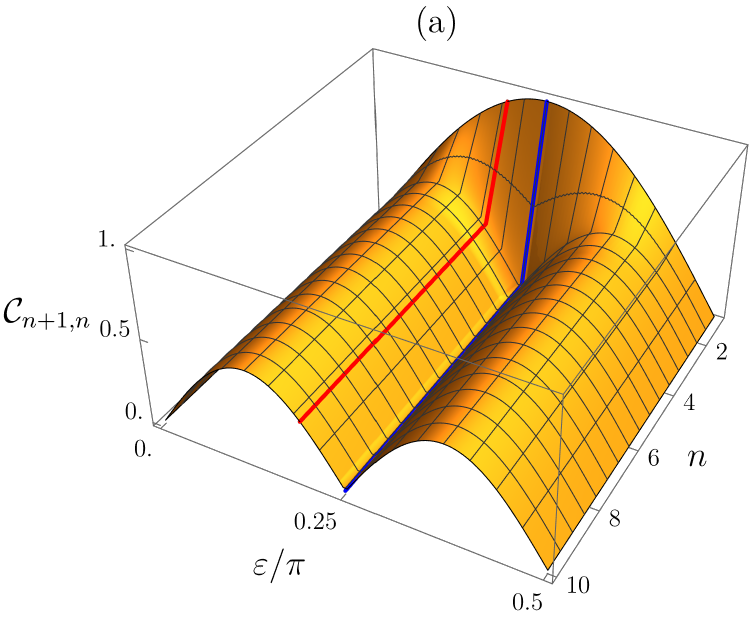}
    \includegraphics[scale=0.65]{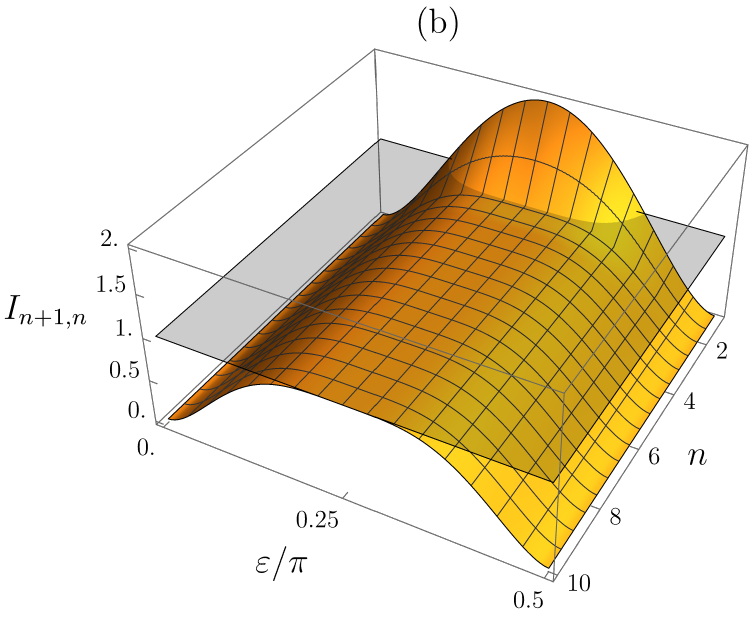}
    \caption{{\bf (a) Concurrence} \eqref{concurrence} {\bf and (b) quantum mutual information} \eqref{qmi} of the state $\rho_{A_nA_{n+1}}$, \eqref{rho_AA}, as a function of the entanglement phase $\varepsilon$ and collision number $n$. The red line in the left panel indicates the case for which the non-Markovianity is maximal, $\varepsilon=\varepsilon_{\rm max}$. When the initial AA entanglement is maximized for $\varepsilon=\frac{\pi}{4}$, indicated by the blue line, the concurrence of $\rho_{A_nA_{n+1}}$ vanishes for $n>1$. The transparent plane in the right panel represents the upper bound on the mutual information for a classically correlated state. For fixed $\varepsilon$, the graphs are constant in $n$ for $n\ge 2$. This is a `boundary effect':  all ancillas $A_n$ with $n\ge 2$ are correlated in the same way with their left and right neighbors in the chain, except for the first ancilla, which is correlated only on one side.}
    \label{fig6}
\end{figure}

In Fig.~\ref{fig6}, we display the behavior of the concurrence with varying entanglement phase $\varepsilon$ after $n-1$ collisions (starting immediately before $S$ collides with $A_1$). Interestingly, we see that $\mathcal{C}(\rho_{A_nA_{n+1}})$ is symmetric about $\varepsilon=\frac{\pi}{4}$ and remains constant for all AA pairs beyond $n=1$. This is in contrast to the non-Markovianity measure in Fig.~\ref{fig5}(a) which is asymmetric over the same interval. In the case of $\varepsilon=\frac{\pi}{4}$, highlighted by the blue line in Fig.~\ref{fig6}(a), the concurrence is maximal for the first AA pair but vanishes for all subsequent pairs. The decoherence function \eqref{Decoresult} for this value of $\varepsilon$ can be evaluated as (see Section \ref{sec6.3})
\begin{equation}
\label{31'}
	\lim_{\varepsilon\rightarrow\pi/4}D(n) = (\cos(2\tau))^n,
\end{equation}
which is strictly nonincreasing in $n$ for all $\tau$. Thus, all system non-Markovianity is {\it eliminated} when the concurrence \eqref{concurrence} of the state $\rho_{A_nA_{n+1}}$ vanishes. We emphasize that the Markovian behavior in this case is qualitatively different to the basic CM where each $A_n-A_{n+1}$ pair are uncorrelated before $A_n$ collides with $S$. To verify this, we examine the quantum mutual information \cite{Nielsen2012} of the state \eqref{rho_AA} in Fig.~\ref{fig6}(b),
\begin{equation}\label{qmi}
I(\rho_{A_nA_{n+1}})=S(\rho_{A_n}) + S(\rho_{A_{n+1}}) - S(\rho_{A_nA_{n+1}}), 
\end{equation}
where $S(\rho)=-\tr[\rho\,{\rm log}\,\rho]$ is the von Neumann entropy of $\rho$. The quantum mutual information $I(\rho_{A_nA_{n+1}})$ provides a measure of the total correlations between $A_n$ and $A_{n+1}$ and is zero only if the two ancillas are in a product state. Apart from the cases $\varepsilon=0$ and $\varepsilon=\frac{\pi}{2}$ corresponding to the basic Markovian setup, we find that ancillas $A_n-A_{n+1}$ are generally correlated for all values of the entanglement phase. In particular, classical correlations between $A_n$ and $A_{n+1}$ are maximized when $\varepsilon=\frac{\pi}{4}$ and $n>1$. This indicates how the transition from non-Markovian to Markovian behavior precisely coincides with the elimination of entanglement between ancillas $A_n$ and $A_{n+1}$ beyond the first interacting pair: as such, we conclude that, within the interacting portion of the environment, AA entanglement is {\it necessary} for the dynamics to be non-Markovian. Moreover, the elimination of AA entanglement resulting from the saturation of entanglement between the first two ancillas may be interpreted as an effect of entanglement monogamy \cite{Coffman2000,Osborne2006}.

To illustrate why non-entangled AA pairs lead to Markovian behavior, let us consider the following upper bound on the trace distance variation between collisions $n$ and $m$ ($m\geq n$) \cite{Laine2010b}
\begin{equation}\label{inequality}
    \mathcal{D}\big(\rho^1_S(m),\rho^2_S(m)\big)-\mathcal{D}\big(\rho^1_S(n),\rho^2_S(n)\big)\leq I_{SA_n}(n),
\end{equation}
with 
\begin{align}\label{bound}
    I_{SA_n}(n) = &\mathcal{D}\big(\rho^1_{SA_n}(n),\rho^1_S(n)\otimes \rho^1_{A_n}(n)\big) + \mathcal{D}\big(\rho^2_{SA_n}(n),\rho^2_S(n)\otimes \rho^2_{A_n}(n)\big) \nonumber\\ 
    &+ \mathcal{D}\big(\rho^1_{A_n}(n),\rho^2_{A_n}(n)\big). 
\end{align}
Here, $\rho_{SA_n}(n)$ represents the joint state of $S-A_n$ immediately {\it before} the $n$-th collision (i.e. after the application of $\mathcal{W}_{[n+1,n]}$), while $\rho_S(n)={\rm Tr}_{A_n}[\rho_{SA_n}(n)]$ and $\rho_{A_n}(n)={\rm Tr}_S[\rho_{SA_n}(n)]$ represent the reduced states of $S$ and the ancilla $A_n$ (as before, the superscripts denote different initial preparations of the system $S$).
The upper bound $I_{SA_n}(n)$ is composed of three terms measuring, respectively, the total correlations between $S$ and $A_{n}$, and the distinguishability of the two reduced states $\rho^1_{A_{n}}$ and $\rho^2_{A_{n}}$. The inequality \eqref{inequality} implies that a nonmonotonic increase in the trace distance can therefore only occur if the system and the interacting portion of ancillas are correlated before the previous $n$-th collision, and/or if the corresponding reduced ancilla states are different. Importantly, this highlights how the emergence of memory effects and information backflow is causally related to the development of correlations between $S$ and $A_{n}$. Based on this analysis, we may introduce the quantities 
\begin{align}
    I_{\rm corr}(n) &= \mathcal{D}\big(\rho^1_{SA_n}(n),\rho^1_S(n)\otimes \rho^1_{A_n}(n)\big) + \mathcal{D}\big(\rho^2_{SA_n}(n),\rho^2_S(n)\otimes \rho^2_{A_n}(n)\big), \label{Icorr}  \\
    I_{\rm env}(n) &= \mathcal{D}\big(\rho^1_{A_n}(n),\rho^2_{A_n}(n)\big),
\end{align}
which measure information contained within system-environment correlations and through the distinguishability of pairs of reduced ancilla states, respectively. We also introduce the measure
\begin{equation}\label{Iext}
    I_{\rm ext}(n) = \mathcal{D}\big(\rho^1_{SA_n}(n),\rho^2_{SA_n}(n)\big) - \mathcal{D}\big(\rho^1_S(n),\rho^2_S(n)\big)
\end{equation}
representing the external information held by ancilla $A_n$ \cite{Breuer2016}.

\begin{figure}[t!]
    \centering
    \includegraphics[scale=0.9]{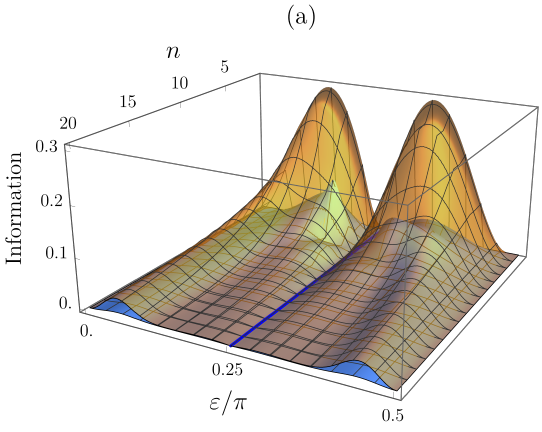}
    \includegraphics[scale=0.9]{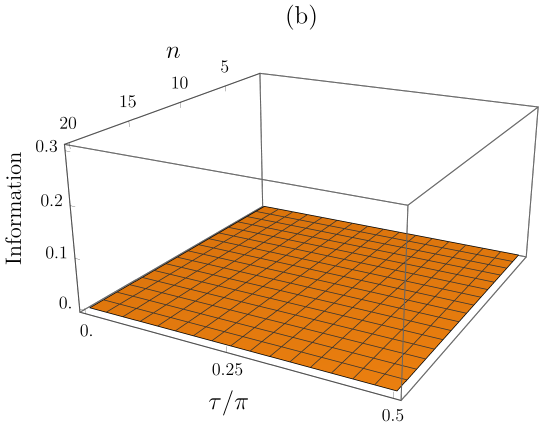}
    \caption{{\bf Information measures $I_{\rm corr}(n)$ (orange surface) and $I_{\rm ext}(n)$ (blue surface)} as a function of the collision number $n$ and {\bf (a)} entanglement phase $\varepsilon$ for $\tau=0.15\pi$, and {\bf (b)} interaction phase $\tau$ for $\varepsilon=\frac{\pi}{4}$. Note that $I_{\rm env}(n)=0$ in both panels since the initial states $\rho^1_{SE}(0)$ and $\rho^2_{SE}(0)$ differ only in the preparation of $\rho^1_S(0)$ and $\rho^2_S(0)$. Each of these system states is parameterized by $p_1=1$, $z_1=0$ and $p_2=0$, $z_2=0$, respectively (see \eqref{dephase_map}). The blue line in {\bf (a)} indicates the $\varepsilon$ value where the AA entanglement vanishes beyond the first interacting pair of ancillas. Along this line both $I_{\rm corr}(n)$ and $I_{\rm ext}(n)$ are zero.}
    \label{fig7}
\end{figure}

In Fig. \ref{fig7}, we display the behavior of these information measures with varying entanglement and interaction phases. Figure \ref{fig7}(a) illustrates the case where the interaction phase is set to $\tau=\tau_{\rm max}$, corresponding to when the non-Markovianity reaches its maximum value for $\varepsilon=\varepsilon_{\rm max}$. Here, the correlation measure $I_{\rm corr}(n)$ is found to be non-zero for almost all values of $\varepsilon$, while $I_{\rm env}(n)$ vanishes due to the initial condition $\rho^1_{E}(0)=\rho^2_{E}(0)$ (recall that $E$ labels the total environment of $n+1$ ancillas).  The non-Markovianity of the dynamics is thus seen to stem from the generation of correlations between $S$ and $A_n$, since the only contribution to the upper bound \eqref{bound} in this case is from $I_{\rm corr}(n)$. The interpretation of non-Markovianity resulting from $S-A_n$ correlations is also consistent with the fact that both $I_{\rm corr}(n)$ and $I_{\rm ext}(n)$ vanish along the line $\varepsilon=\frac{\pi}{4}$, coinciding with the loss of AA entanglement and the transition from non-Markovian to Markovian dynamics (see Figs. \ref{fig5}(a) and \ref{fig6}(a)). In particular, the same quantities are shown in Fig. \ref{fig7}(b) with varying $\tau$ and fixed entanglement phase $\varepsilon=\frac{\pi}{4}$, where it is found that all three measures vanish. This demonstrates how the transition to Markovian behavior is caused by the loss of correlations between the system $S$ and ancilla $A_n$, which occurs in parallel to the loss of AA entanglement between the interacting portion of ancillas.

Finally, we note that for $\tau=\frac{\pi}{4}$, the decoherence function $D(n)$, \eqref{Decoresult} satisfies the following limit (proven in Section \ref{sec6.3}):
\begin{equation}
\label{32'}    
\lim_{\tau\rightarrow\pi/4}D(n) = e^{-i\pi n/2}\cos(2\varepsilon),\qquad n\geq 1.
\end{equation}
The system behavior is therefore Markovian in this case regardless of the value of $\varepsilon$, given that $|D(n)|$ reaches its steady state value after the first collision. This is consistent with the numerical results shown in Figs. \ref{fig4} and \ref{fig5}(b), in which $\mathcal{N}_{BLP}$ vanishes for $\tau=\frac{\pi}{4}$.

\subsection{Interchanging the order of SA and AA collisions}
\label{ssec:main}

As we have seen in the previous section, AA entanglement is necessary for the system evolution \eqref{m21.1'} to exhibit memory effects. We further show this occurrence of non-Markovianity is due to the fact that the ancillas are entangled {\it before} they collide with the system.

For this, we proceed to evaluate the dynamical map \eqref{m46'} in the opposite case, where consecutive ancillas are entangled {\it after} the first member in each pair collides with $\s$. In particular, this coincides with the CCM representation of the open dynamics depicted in Fig.~\ref{fig3}(b). The dynamical map in this case reads 
\begin{equation}
\label{118m'}
{\M}'[\vartheta] = \tr_{\a_1}\big(\e^{-i\varepsilon\bbbone\otimes\sigma^{(2)}_z\otimes\sigma^{(1)}_z}\left(\left[\e^{-i\tau\sigma_x\otimes\sigma^{(1)}_x}\vartheta\e^{i\tau\sigma_x\otimes\sigma^{(1)}_x}\right]\otimes\rho_\a\right)\e^{i\varepsilon\bbbone \otimes\sigma^{(2)}_z\otimes\sigma^{(1)}_z}\big).
\end{equation}
We show in Sec. \ref{sec:interchange} that the corresponding decoherence function is given by
\begin{equation}
\label{d'n}
D'(n)= \e^{-2i\tau} \big(\cos(2\tau) -i\sin(2\tau)\cos(2\varepsilon)\big)^{n-1}, \quad n\ge 1.
\end{equation}
Hence, $|D'(n)|=(\cos^2(2\tau)+\sin^2(2\tau)\cos^2(2\varepsilon))^{\frac{1}{2}(n-1)}$, which is a function of $n$ that never increases, since $\cos^2(2\tau)+\sin^2(2\tau)\cos^2(2\varepsilon)\le 1$. We conclude that the system dynamics is Markovian for all values of the parameters $\tau$ and $\varepsilon$. This illustrates how reversing the order of SA and AA operations can profoundly affect the non-Markovianity of the system evolution.

\section{Conclusions}
\label{sec5}

In summary, we have derived a new class of non-Markovian CMs with correlated environment states, whose correlations are generated through applying a sequence of unitary (or CPTP) operations to successive groups of ancillas. For these types of CMs, we have demonstrated that the open dynamics admits a general mapping onto a CCM, in which the memory part of the environment is incorporated into an enlarged Markovian system. In particular, the size of this memory part is quantified by the correlation length characterizing the range of intra-environment correlations between ancillas. This shares an analogous interpretation with the memory depth from Ref. \cite{Campbell2018}, which, for a separate class of CMs featuring AA collisions, quantifies the range of system-environment correlations that must be accounted for in the construction of the system dynamical map. Notably, the CCM corresponding to our setup displays key differences to that of \cite{Campbell2018} regarding the arrangement of the memory and non-memory parts and their mutual interactions with the system.

We have further analyzed the extent to which entanglement can influence the non-Marko\-vianity of the open dynamics of an all-qubit CM with correlated ancilla pairs. By examining the nature of correlations shared between ancilla pairs representing the interacting portion of the environment, we have found AA entanglement to be necessary (although not sufficient) for the emergence of non-Markovian behavior. At the same time, the order in which SA and AA collisions occur has been shown to play a key role in determining the non-Markovianity of the open dynamics. A more detailed analysis of the role of AA entanglement in the emergence of non-Markovianity will be presented elsewhere \cite{PleasanceTBC}.

Looking ahead, the CM developed here could be used to study the thermodynamics of open systems interacting with nonthermal environments, whereby ancillas initially prepared in thermal states are correlated in sequence. Understanding how such correlations impact the performance of thermal devices---taking into account the energetic cost in generating such correlations---may have applications in the design of quantum heat engines and refrigerators \cite{DeChiara2020}. On the experimental side, our model could be implemented via the same platforms used to realize CMs featuring AA collisions, including NMR \cite{Bernardes2016} and optical setups \cite{Jin2015, Cuevas2019}, or it could also be simulated on a quantum computer \cite{Garcia2020}. Finally, one could also extend our work in the direction of Refs. \cite{Lorenzo2017,Rybar2012,Filippov2017} by formally characterizing the non-Markovian dynamics simulable by a CM with a correlation structure determined by \eqref{m12}. For an open two-level system, it has been demonstrated that any CP- or P-indivisible Pauli channel can be simulated by a CM with suitably correlated ancillas \cite{Rybar2012,Filippov2017}. While the precise connection between non-Markovianity and intra-environment correlations is not yet fully understood, such work could shed further light on the applicability of CMs for simulating general non-Markovian open dynamics.

\section{Proof of results}
\subsection{Proof of Theorem \ref{thm1}}
\label{deriv:21.1}

Starting from \eqref{reduced_state},
\begin{equation}
\rho_\s(n) = \tr_{\a_{n+L-1},\ldots,\a_{1}} \big[ \tr_{\a_n}\U_{\s n} \W_{[n,n-L+1]}\cdots \tr_{\a_1}\U_{\s 1}\W_{[L,1]} (\rho_\s\otimes\rho^{\otimes n+L-1}_\a) \big],
\label{m15}
\end{equation}
we define the map $\wM : \mathcal{S}(\h_S\otimes\h^{\otimes L}_A)\rightarrow \mathcal{S}(\h_S\otimes\h^{\otimes L}_A)$ acting on density matrices $\varrho$ of $\s$ and $L$ ancillas by 
\begin{equation}
\label{m16}
\wM[\varrho] = \Big( \tr_{\a_1} \U_{\s 1}\W_{[L,1]}\varrho\Big)\otimes\rho_\a.
\end{equation}
Here, the trace is taken over the first ancilla space (the one that $\s$ interacts with via $\U_{\s 1}$).  The operator $\wM[\varrho]$ is again a density matrix of the system plus $L$ ancillas, with the first ancilla traced out and replaced by a fresh one in the state $\rho_A$. Writing $\wM^n =\wM\circ\cdots\circ \wM$, we obtain from \eqref{m15},
\begin{equation}
\label{m17}
\rho_\s(n) = \tr_{\a_L,\ldots,\a_1} \wM^n[\rho_\s\otimes \rho^{\otimes L}_\a].
\end{equation}
Now let $\vartheta_0$ be a density matrix of the system $\s$ plus $L-1$ ancillas. According to \eqref{m16}, we have $\wM[\vartheta_0\otimes\rho_\a] = \vartheta_1\otimes\rho_\a$, where $\vartheta_1$ is again a density matrix of $\s$ plus $L-1$ ancillas. The definition of the map $\M$ from \eqref{20m} implies $\mathcal M[\vartheta_0]=\vartheta_1$, so that $\wM[\vartheta_0\otimes\rho_\a ] =\mathcal M[\vartheta_0]\otimes\rho_\a$, and
\begin{equation}
\wM^n[\vartheta_0\otimes\rho_\a] = \mathcal M^n[\vartheta_0]\otimes\rho_\a.
\end{equation}
Combining this with \eqref{m17} finally provides \eqref{m21.1} of the main text. 

It is straightforward to further prove the CPTP property of $\M$ by noting that 
\begin{equation*}
    \M[\vartheta] = \tr_{\a_L}\Big(\mathcal S_{L,1}\U_{\s 1}\W_{[L,1]} (\vartheta\otimes\rho_\a)\Big) = \tr_{\a_L}\big(U(\vartheta\otimes\rho_\a)U^{\dagger}\big).
\end{equation*}
Here, $\mathcal{S}_{L,1}$ is a swap map that switches the first and last ancillas in the tensor product $\h_S\otimes\h^{\otimes L}_A$, and $U$ is a fixed unitary operator. It directly follows from Stinespring's theorem \cite{Stinespring1955} that the resulting map $\vartheta\mapsto \tr_{\a_L}\big(U(\vartheta\otimes\rho_\a)U^{\dagger}\big)$ is CPTP. This completes the proof of Theorem \ref{thm1}. \hfill $\square$ 

\subsection{Proof of Proposition \ref{prop1-1}}
\label{sec:DerivAnfin}

We first note that the swap map $\mathcal{S}_{k,l}$ acting on any ancilla pair $A_k$ and $A_l$ obeys the identities
\begin{equation}
    \mathcal S_{k,l}^2=\mathcal{I}, \qquad \U_{\s 1}\mathcal S_{2,1}\mathcal S_{3,2}\cdots \mathcal S_{L,L-1}=\mathcal S_{2,1}\mathcal S_{3,2}\cdots \mathcal S_{L,L-1}\U_{\s L},
    \label{swap_idents}
\end{equation}
where $\mathcal{I}$ denotes the identity map on $\h_A\otimes\h_A$. The second identity relates to the fact that transferring the state of $A_L$ to $A_1$, colliding $S$ with $A_1$, and then shifting  the state of $A_1$ back to $A_L$, is in effect the same as colliding $S$ with $A_L$. Using the explicit form of $\M$ as per \eqref{20m},
\begin{equation*}
    \M[\vartheta] = \tr_{A_1}\Big(\U_{\s 1}\W_{[L,1]} (\vartheta\otimes\rho_\a)\Big),
\end{equation*}
we apply \eqref{swap_idents} to obtain 
\begin{eqnarray}
\mathcal M[\vartheta] &=&\tr_{\a_1}\Big( \U_{\s 1} \mathcal S_{2,1}\mathcal S_{3,2}\cdots\mathcal S_{L,L-1}\mathcal S_{L,L-1}\cdots\mathcal S_{3,2}\mathcal S_{2,1}\W_{[L,1]} (\vartheta\otimes\rho_\a)\Big) \nonumber\\
    &=&\tr_{\a_1}\Big(\mathcal S_{2,1}\mathcal S_{3,2}\cdots \mathcal S_{L,L-1}\U_{\s L}\W_{[L,1]}' (\vartheta\otimes\rho_\a)\Big),
\label{an1}
\end{eqnarray}
with $\mathcal{W}'_{[L,1]}=\mathcal{S}_{L,L-1}\cdots\mathcal{S}_{2,1}\mathcal{W}_{[L,1]}$. Next we use that
\begin{equation}
    \tr_{\a_k}\Big(\mathcal S_{k+1,k}\, \rho_\s\rho_{\a_1}\cdots\rho_{\a_k}\rho_{\a_{k+1}}\cdots \Big)=\rho_\s\cdots\rho_{\a_k}\rho_{\a_{k+2}}\cdots=\tr_{\a_{k+1}}\Big(\rho_\s\cdots\rho_{\a_k}\rho_{\a_{k+1}}\cdots\Big).
\end{equation}
By linearity and applying this relation $L-1$ times to the last line of \eqref{an1}, we then get \eqref{anfin}. This completes the proof of Proposition \ref{prop1-1}. \hfill $\square$

\subsection{Proof of Proposition \ref{prop2}}
\label{proofprop2}

For the model outlined in Sec. \ref{sec4}, in which both $S$ and each $A$ are qubits, we may express $\M$ in the following ordered operator basis
\begin{equation}
\label{m47}
\big\{ e_{00}, e_{01}, e_{02}, e_{03}, \quad
 e_{10}, e_{11}, e_{12}, e_{13},\quad  e_{20}, e_{21}, e_{22}, e_{23}, \quad 
 e_{30}, e_{31}, e_{32}, e_{33} \big\},
\end{equation}
where
\begin{equation}
e_{kl} = e_k\otimes e_l,
\end{equation}
and
\begin{equation}
e_0 = \frac{1}{\sqrt 2}\bbbone,\quad
e_1 = \frac{1}{\sqrt 2}\sigma_x,\quad
e_2 = -\frac{i}{\sqrt 2}\sigma_y,\quad
e_3 = \frac{1}{\sqrt 2}\sigma_z.
\end{equation}
(To simplify the notation we shall use the same $e_k$ to denote basis operators of either $S$ or $A$.) The operator sets $\{e_j\}$ are mutually orthonormal with respect to the Hilbert-Schmidt inner product 
\begin{equation}
\langle e_k | e_l\rangle \equiv \tr[e^{\dagger}_k e_l] = \delta_{k,l},
\label{m48}
\end{equation}
such that 
\begin{equation}
\langle e_{kl} | e_{mn}\rangle = \langle e_k\otimes e_l | e_m\otimes e_n\rangle = \langle e_k | e_m\rangle\, \langle e_l | e_n\rangle = \delta_{k,m}\delta_{l,n}.
\end{equation} 
Accordingly, when expressed in the basis \eqref{m47}, the map $\mathcal M$ takes the form of a $16\times16$ matrix with matrix elements 
\begin{equation*}
\boldsymbol{M}_{kl,mn} \equiv \langle e_{kl}|\boldsymbol{M}|e_{lm}\rangle = \tr[(e_k\otimes e_l)^{\dagger}\M[e_m\otimes e_n]] , 
\end{equation*}
while in the same basis, the composite state $\vartheta\mapsto|\vartheta\rangle$ of $S+A$ is represented as $16\times 1$ column vector. The coefficients $\rho_m$ parameterizing $\rho_A$ may also be computed from the inner product \eqref{m48} as  
\begin{equation}
\label{comp}
\rho_m = \sqrt 2 \langle e_m|\rho_\a\rangle = \sqrt2 \tr[e_m^{\dagger} \rho_\a],
\end{equation}
with $\rho_0=1$.

After much algebra, the matrix $\boldsymbol{M}$ can be shown to take the form 
\begin{equation}
\label{m56}
\boldsymbol{M} = \begin{pNiceArray}{c|c|c|c}
\boldsymbol{A}_1 & 0 & & \\
\cline{1-2}
0 & \boldsymbol{A}_1 & & \\
\hline
 & &  \cos(2\tau) \boldsymbol{A}_1 & i\sin(2\tau) \boldsymbol{A}_2\\
 \cline{3-4}
 & & i\sin(2\tau) \boldsymbol{A}_2 & \cos(2\tau) \boldsymbol{A}_1
\end{pNiceArray}.
\end{equation}
Here, the empty blocks have null entries, and the matrices $\boldsymbol{A}_1$ and $\boldsymbol{A}_2$ are given by
\begin{equation}
\boldsymbol{A}_1 =
\begin{pmatrix}
1 & 0 & 0 & 0\\
\rho_1\cos(2\varepsilon) & 0 & 0 &-i\rho_2\sin(2\varepsilon)\\
\rho_2\cos(2\varepsilon) & 0 & 0 & -i\rho_1\sin(2\varepsilon)\\
\rho_3 & 0 & 0 & 0
\end{pmatrix},\qquad 
\boldsymbol{A}_2 =
\begin{pmatrix}
0 & \cos(2\varepsilon) & -i\rho_3\sin(2\varepsilon) & 0\\
0 & \rho_1 & 0 & 0\\
0 & \rho_2 & 0 & 0\\
0 & \rho_3\cos(2\varepsilon) & -i\sin(2\varepsilon) & 0 
\end{pmatrix}.
\label{A1A2appen}
\end{equation}
The structure of \eqref{m56} into the $4\times 4$ blocks corresponds to the four groupings of the ordered basis indicated by a space in \eqref{m47}.

The matrix $\boldsymbol{A}_1$ has a simple eigenvalue $1$, and only one other eigenvalue $0$ of algebraic multiplicity $3$ and geometric multiplicity $2$, implying it is nondiagonalizable. However, powers of $\boldsymbol{A}_1$ are straightforward to calculate,
\begin{equation}
\label{m58}
\boldsymbol{A}_1^n = 
\begin{pmatrix}
1 & 0 & 0 & 0\\
\rho_1\cos(2\varepsilon) -i\rho_2\rho_3\sin(2\varepsilon) & 0 & 0 &0\\
\rho_2\cos(2\varepsilon) -i \rho_1\rho_3\sin(2\varepsilon) & 0 & 0 & 0\\
\rho_3 & 0 & 0 & 0
\end{pmatrix}
\qquad n\ge 2.
\end{equation}
Hence, the $n$-th power of \eqref{m56} is then
\begin{equation}
\label{m59}
\boldsymbol{M}^n = \begin{pNiceArray}{c|c|c}
\boldsymbol{A}_1^2 & 0 &  \\
\cline{1-2}
0 & \boldsymbol{A}_1^2 &  \\
\hline
 & &  \boldsymbol{B}^n
\end{pNiceArray}, 
\qquad n\ge 2,
\end{equation}
where the $8\times 8$ matrix $\boldsymbol{B}$ is defined as
\begin{equation}
\label{mB}
\boldsymbol{B} = \begin{pNiceArray}{c|c}
\cos(2\tau) \boldsymbol{A}_1 & i\sin(2\tau) \boldsymbol{A}_2\\
\hline
  i\sin(2\tau) \boldsymbol{A}_2 & \cos(2\tau) \boldsymbol{A}_1
\end{pNiceArray}.
\end{equation}

Using the above expression for $\boldsymbol{M}^n$, we next show that the decoherence function $D(n)$ can be evaluated in terms of powers of the matrices $\boldsymbol{A}_{1,2}$, \eqref{Decoresult}. 

\begin{lem}
\label{lem:1}
The decoherence function $D(n)$ in \eqref{dephase_map} is given by
\begin{equation}\label{m45}
D(n) = \sum_{m=0}^3([\boldsymbol{M}^n]_{30,3m} - [\boldsymbol{M}^n]_{20,3m})\rho_m.
\end{equation}
\end{lem}
\noindent {\it Proof.} In the representation introduced above, we can evaluate the reduced density matrix $\rho_S(n)=\tr_\a \mathcal M^n [\rho_\s\otimes\rho_\a]$ by acting $\boldsymbol{M}^n$ on the column vector $|\rho_S\otimes\rho_A\rangle$. Since $\tr_\a [e_{kl}]=\sqrt{2} e_k\delta_{l,0}$, the parts of the image of $\boldsymbol{M}^n$ proportional to $e_{kl}$ all vanish except if $l=0$, and hence we do not need all matrix elements of $\boldsymbol{M}^n$ to determine $\rho_S(n)$. Explicitly, we have
\begin{eqnarray}
\rho_S(n) &=& \sum_{k,l=0}^3 \tr_\a[e_{kl}]\langle e_{kl}| \boldsymbol{M}^n |\rho_\s\otimes\rho_\a\rangle \nonumber\\
 &=& \sqrt{2} \sum_{k=0}^3 e_k \langle e_{k0}| \boldsymbol{M}^n |\rho_\s\otimes\rho_\a\rangle\nonumber\\
 &=& \sqrt{2} \sum_{k=0}^3 e_k \sum_{l,m=0}^3[\boldsymbol{M}^n]_{k0,lm}\langle e_{lm}|\rho_\s\otimes\rho_\a\rangle.
 \label{m72}
\end{eqnarray}
Next we have from \eqref{comp},
\begin{equation*}
\langle e_{lm} |\rho_\s\otimes\rho_\a\rangle =
\frac{1}{\sqrt{2}}\rho_m \tr[e_l^{\dagger}\rho_\s],\qquad l,m=0,\ldots,3, 
\end{equation*}
which combined with \eqref{m72} yields
\begin{equation}
\label{m73}
\rho_S(n) = \sum_{k=0}^3 e_k \sum_{l,m=0}^3[\boldsymbol{M}^n]_{k0,lm}\rho_m  \tr[e_l^{\dagger}\rho_\s].
\end{equation}
We can further simplify this expression using (see \eqref{m59}, \eqref{m56}) 
\begin{equation}
[\boldsymbol{M}^n]_{00,lm} = \delta_{l,0}\delta_{m,0}, \qquad [\boldsymbol{M}^n]_{10,lm} = \delta_{l,1}\delta_{m,0},
\end{equation}
so that
\begin{equation}
\label{m75}
\rho_S(n) = \frac{1}{\sqrt 2} e_0  + e_1 \tr[e_1\rho_\s] +  \sum_{k=2,3} e_k \sum_{l,m=0}^3 [\boldsymbol{M}^n]_{k0,lm}\rho_m  \tr[e_l^{\dagger}\rho_\s].
\end{equation}
The initial state $\rho_S$ represented in the $\sigma_x$ basis reads 
\begin{equation}
\label{rhox}
\rho_\s = 
\begin{pmatrix}
p & z\\
z^* & 1-p
\end{pmatrix},\qquad (\sigma_x{\rm-basis})
\end{equation}
with $p\in [0,1]$ and $|z|^2\le p(1-p)$. Let us then also represent the reduced density matrix \eqref{m75} in the same basis,
\begin{equation}
\label{m76}
\rho_\s(n) =
\begin{pmatrix}
p & \alpha_3(n) -\alpha_2(n)\\
\alpha_3(n)+\alpha_2(n) & 1-p
\end{pmatrix},\qquad (\sigma_x{\rm-basis})
\end{equation}
where
\begin{equation}
\label{m78}
\alpha_k(n) = \frac{1}{\sqrt2}\sum_{l,m=0}^3[\boldsymbol{M}^n]_{k0,lm}\rho_m  \tr[e_l^{\dagger}\rho_\s], \quad k=2,3.
\end{equation}
In terms of the parameters $p$ and $z$, we have 
\begin{equation}
\label{m79}
\tr[e_l^{\dagger}\rho_\s] =\sqrt 2\left\{
\begin{array}{ll}
\tfrac12, & l=0\\
p-\tfrac12,& l=1\\
-i{\rm Im}\,z, & l=2\\
{\rm Re}\,z, & l=3
\end{array}
\right.
\end{equation}
The reduced system dynamics is thus entirely determined by the off-diagonal matrix elements of \eqref{m76} depending on $\alpha_2(n)$ and $\alpha_3(n)$ (in the $\sigma_x$ basis). We now define 
\begin{equation}
\label{dnz}
\alpha_3(n)-\alpha_2(n) = D(n) z,
\end{equation}
where $D(n)$ is a function of $n$ which is independent of $z$. In particular, both off-diagonal elements of $\rho_S(n)$ must vanish for all $n$ if $z=0$. It then follows that $\alpha_2(n)=\alpha_3(n)=0$ if $z=0$. From \eqref{m56} and \eqref{m79}, the terms with $l=0,1$ in \eqref{m78} can also be seen to vanish, and so
\begin{equation}
\alpha_k(n) = \frac{1}{\sqrt2} \sum_{l=2,3} \ \sum_{m=0}^3[\boldsymbol{M}^n]_{k0,lm}\rho_m  \tr[e_l^{\dagger}\rho_\s], \quad k=2,3.
\end{equation}
Using again \eqref{m79} we obtain
\begin{equation}
\alpha_k(n) = \bigg(\sum_{m=0}^3[\boldsymbol{M}^n]_{k0,3m}\rho_m \bigg) {\rm Re}\,z -i \bigg( \sum_{m=0}^3[\boldsymbol{M}^n]_{k0,2m} \rho_m \bigg) {\rm Im }\,z, \quad k=2,3.
\end{equation}
For the left side of \eqref{dnz} to be a multiple of $z$, we have the constraint
\begin{equation}
\sum_{m=0}^3([\boldsymbol{M}^n]_{30,3m} - [\boldsymbol{M}^n]_{20,3m})\rho_m = -\sum_{m=0}^3([\boldsymbol{M}^n]_{30,2m} -[\boldsymbol{M}^n]_{20,2m})\rho_m
\end{equation}
and the decoherence function $D(n)$ becomes \eqref{m45}. This completes the proof of Lemma \ref{lem:1}.\hfill $\square$

\smallskip

Next we evaluate the matrix elements appearing in the sum \eqref{m45}. We first obtain from \eqref{m56},
{\begin{equation}\label{decoherencB}
D(n)=\sum_{m=0}^3([\boldsymbol{B}^n]_{4,m+4} - [\boldsymbol{B}^n]_{0,m+4})\rho_m,
\end{equation}
where $\boldsymbol{B}$ is given in \eqref{mB}. We observe that $\boldsymbol{B}$ is a block circulant matrix and that it can be block-diagonalized according to
\begin{equation}\label{equations}
\boldsymbol{B} = \frac{1}{2}\left(\begin{array}{cc}
\bbbone_4 & \bbbone_4 \\
\bbbone_4 & -\bbbone_4
\end{array}\right)\left(\begin{array}{cc}
\boldsymbol{\Xi}_{+} & 0\\
0 & \boldsymbol{\Xi}_{-}
\end{array}\right)\left(\begin{array}{cc}
\bbbone_4 & \bbbone_4 \\
\bbbone_4 & -\bbbone_4
\end{array}\right)
\end{equation}	
where $\bbbone_4$ are $4\times 4$ identity matrices, and 
\begin{equation}
\boldsymbol{\Xi}_{\pm}=\cos(2\tau) \boldsymbol{A}_1 \pm i\sin(2\tau) \boldsymbol{A}_2.
\end{equation}
Let us denote the Jordan decomposition of $\boldsymbol{\Xi}_\pm$ by
\begin{equation}
\label{decomposition}
\boldsymbol{\Xi}_{\pm} = \boldsymbol{P}_{\pm}\boldsymbol{D}_{\pm}\boldsymbol{P}_{\pm}^{-1},
\end{equation}
where $\boldsymbol{P}_\pm$ are invertible matrices and $\boldsymbol{D}_{\pm}$ are the Jordan blocks. It follows from \eqref{equations} and \eqref{decomposition} that 
\begin{eqnarray}
\boldsymbol{B}^n &=& \frac{1}{2}\left(\begin{array}{cc}
\boldsymbol{P}_+ & \boldsymbol{P}_- \\
\boldsymbol{P}_+ & -\boldsymbol{P}_-
\end{array}\right)\left(\begin{array}{cc}
\boldsymbol{D}_{+}^n & 0\\
0 & \boldsymbol{D}_{-}^n
\end{array}\right)\left(\begin{array}{cc}
\boldsymbol{P}_+^{-1} & \boldsymbol{P}_+^{-1} \\
\boldsymbol{P}_-^{-1} & -\boldsymbol{P}_-^{-1}
\end{array}\right)\nonumber\\
&=&\frac{1}{2}\left(\begin{array}{cc}
\boldsymbol{P}_+\boldsymbol{D}_{+}^n\boldsymbol{P}_+^{-1}+\boldsymbol{P}_-\boldsymbol{D}_+^n\boldsymbol{P}_-^{-1} & \boldsymbol{P}_+\boldsymbol{D}_+^n\boldsymbol{P}_+^{-1}-\boldsymbol{P}_-\boldsymbol{D}_+^n\boldsymbol{P}_-^{-1}\\
\boldsymbol{P}_+\boldsymbol{D}_+^n\boldsymbol{P}_+^{-1}-\boldsymbol{P}_-\boldsymbol{D}_+^n\boldsymbol{P}_-^{-1} & \boldsymbol{P}_+\boldsymbol{D}_+^n\boldsymbol{P}_+^{-1}+\boldsymbol{P}_-\boldsymbol{D}_+^n\boldsymbol{P}_-^{-1}
\end{array}\right).
\label{BnPDP}
\end{eqnarray}
In view of \eqref{decoherencB}, we use \eqref{BnPDP} to get 
\begin{equation}
\boldsymbol{B}_{0,m+4}=\frac{1}{2}[\boldsymbol{P}_+\boldsymbol{D}_+^n\boldsymbol{P}_+^{-1}]_{0m}-\frac{1}{2}[\boldsymbol{P}_-\boldsymbol{D}_-^n\boldsymbol{P}_-^{-1}]_{0m}
\end{equation}
and
\begin{equation}
\boldsymbol{B}_{4,m+4}=\frac{1}{2}[\boldsymbol{P}_+\boldsymbol{D}_+^n\boldsymbol{P}_+^{-1}]_{0m}+\frac{1}{2}[\boldsymbol{P}_-\boldsymbol{D}_-^n\boldsymbol{P}_-^{-1}]_{0m}.
\end{equation}
Substituting the above into \eqref{decoherencB} leads to 
\begin{equation}
\label{Decoresult2}
D(n)=\sum_{m=0}^3[\boldsymbol{P}_-\boldsymbol{D}_-^n\boldsymbol{P}_-^{-1}]_{0m}\rho_m=\sum_{m=0}^3[\boldsymbol{\Xi}_-^n]_{0m}\rho_m,
\end{equation}
which gives precisely \eqref{Decoresult}. This concludes the proof of Proposition \ref{prop2}. \hfill $\square$

\subsection{Derivation of \eqref{31'} and \eqref{32'}}
\label{sec6.3}

In this section, we derive limiting cases of the decoherence function \eqref{Decoresult2}, assuming each ancilla is initialized in the state $\rho_A=|+\rangle\langle+|_A$, \eqref{rhoa}. Since for this state $\rho_1=1$ and $\rho_2=\rho_3=0$, from \eqref{A1A2appen} we have
\begin{eqnarray}
\label{an86}
\lefteqn{\cos(2\tau)\boldsymbol{A}_1-i\sin(2\tau)\boldsymbol{A}_2}\nonumber \\
&=&
\begin{pmatrix}
\cos(2\tau) & -i\sin(2\tau)\cos(2\varepsilon) & 0 & 0 \\
\cos(2\tau)\cos(2\varepsilon) & -i\sin(2\tau) & 0 & 0 \\
0 & 0 & 0 & -i\cos(2\tau)\sin(2\varepsilon)  \\
0 & 0 & -\sin(2\tau)\sin(2\varepsilon) & 0 
\end{pmatrix},
\end{eqnarray}
such that combining \eqref{Decoresult2} and \eqref{an86} yields
\begin{equation}
D(n)=\left[\boldsymbol{C}^n\right]_{00}+\left[\boldsymbol{C}^n\right]_{01},\qquad \boldsymbol{C} =
\begin{pmatrix}
\cos(2\tau) & -i\sin(2\tau)\cos(2\varepsilon) \\
\cos(2\tau)\cos(2\varepsilon) & -i\sin(2\tau)
\end{pmatrix}.
\label{an87}
\end{equation}
The matrix $\boldsymbol{C}$ can be diagonalized as
\begin{equation}
\label{88}
\boldsymbol{C} = \boldsymbol{P}
\begin{pmatrix}
z_+ & 0 \\
0 & z_-
\end{pmatrix}\boldsymbol{P}^{-1},
\end{equation}
where the eigenvalues are $z_\pm = \frac12\big(e^{-2i\tau} \pm \Delta\big)$, and $\Delta=\sqrt{\cos(4\tau)-i\cos(4\varepsilon)\sin(4\tau)}$. Furthermore, the matrix $\boldsymbol{P}$ and its inverse $\boldsymbol{P}^{-1}$
read 
\begin{equation}
\boldsymbol{P}=
\begin{pmatrix}
q_+ & q_-  \\
1 & 1
\end{pmatrix},
\qquad \boldsymbol{P}^{-1} = \frac{1}{\kappa}
\begin{pmatrix}
1 \  & -q_-  \\
-1\  & q_+
\end{pmatrix},
\end{equation}
with
\begin{equation}
q_\pm(\varepsilon,\tau) = \frac{\kappa}{2\Delta}(e^{2i\tau}\pm \Delta), \qquad \kappa(\varepsilon,\tau)=q_+-q_-. 
\end{equation}
Let us for now assume that $\varepsilon,\tau\neq\frac{\pi}{4}$, implying $\cos(2\varepsilon)\cos(2\tau)\neq 0$. We can then proceed to calculate powers of $\boldsymbol{C}$ using \eqref{88}. In doing so, the decoherence function \eqref{an87} becomes $D(n) = \frac{1}{\kappa} ( q_+(1-q_-) z_+^n-q_-(1-q_+)z_-^n)$, which can be rewritten as
\begin{equation}
D(n) =\zeta_+(\varepsilon,\tau)\left(\dfrac{\e^{-2i\tau}+\Delta}{2}\right)^n- \zeta_-(\varepsilon,\tau)\left(\dfrac{\e^{-2i\tau}-\Delta}{2}\right)^n,
\label{dec1}
\end{equation}
where 
\begin{equation}
\label{91}
\zeta_\pm(\varepsilon,\tau) = \frac{1}{\Delta} \Big(1-\frac{e^{2i\tau}\mp \Delta}{2\cos(2\varepsilon)\cos(2\tau)}\Big) \frac{e^{2i\tau}\pm\Delta}{2}.
\end{equation}
Note that since $|\Delta|^4=\cos^2(4\tau)+\sin^2(4\tau)\cos^2(4\varepsilon)\le 1$, with equality if and only if $\cos(4\varepsilon)=\pm1$, we have $|\Delta|<1$ for $\varepsilon\neq\frac{\pi}{4}$, and therefore $\frac12|e^{-2i\tau}\pm\Delta|<1$. It follows from \eqref{dec1} that $|D(n)|\le Ce^{-\gamma n}$ for some $C,\gamma>0$. This implies initial coherences of $S$ to decay exponentially in the collision number $n$ whenever $\varepsilon\neq\frac{\pi}{4}$. 

Next we consider the following two limits of \eqref{dec1}, which are well defined given that $D(n)$ is a continuous function of both $\varepsilon$ and $\tau$:
\begin{itemize}
    \item[(i)] As $\varepsilon\rightarrow\frac{\pi}{4}$, we obtain $\zeta_+(\pi/4,\tau)=1$ and $\zeta_-(\pi/4,\tau)=0$, giving
    \begin{equation}
 \label{f92}   \lim_{\varepsilon\rightarrow\pi/4}D(n) = (\cos{(2\tau)})^n.
\end{equation}
\item[(ii)] As $\tau\rightarrow\frac{\pi}{4}$, we obtain for $n\geq1$,
\begin{equation*}
    \lim_{\tau\rightarrow\pi/4}\zeta_+(\varepsilon,\tau)\left(\frac{\e^{-2i\tau}+\Delta}{2}\right)^n=0, \quad 
    \lim_{\tau\rightarrow\pi/4}\zeta_-(\varepsilon,\tau)\left(\frac{\e^{-2i\tau}-\Delta}{2}\right)^n=-(-i)^n\cos({2\varepsilon}),
\end{equation*}
such that
\begin{equation}
\label{f93}
\lim_{\tau\rightarrow\pi/4} D(n) = e^{-i\pi n/2}\cos(2\varepsilon)\qquad  n\ge 1.
\end{equation}
\end{itemize}
Hence, we recover the two expressions \eqref{31'} and \eqref{32'}. 

\subsection{Derivation of \eqref{d'n}}
\label{sec:interchange}

The decoherence function $D'(n)$ can be derived following the same procedure used to obtain $D(n)$ in Sec. \ref{proofprop2}, where $\M'$ can be similarly represented as a $16\times 16$ matrix. For the sake of brevity, we omit the full derivation and only present the intermediate results needed to construct the solution for $D'(n)$. The matrix representation of \eqref{118m'} in the basis \eqref{m47} is found to be
\begin{equation}
\boldsymbol{M}' = 
\begin{pNiceArray}{cc|cc}
\boldsymbol{A}_1' & \boldsymbol{A}_2' &  &  \\
\boldsymbol{A}_2' & \boldsymbol{A}_1' &  &  \\ 
\hline
&  & \boldsymbol{B}_1 & \boldsymbol{B}_2 \\
&  & \boldsymbol{B}_2 & \boldsymbol{B}_1
\end{pNiceArray},
\end{equation}
where 
\begin{equation}
\boldsymbol{A}_1'=\boldsymbol{A}_1,\quad 
\boldsymbol{A}_2'=
\begin{pmatrix}
0 & 0 & 0 & 0 \\
0 & 0 & -i\rho_2\sin{(2\varepsilon)} & 0 \\
0 & 0 & -i\rho_1\sin{(2\varepsilon)} & 0 \\
0 & 0 & 0 & 0
\end{pmatrix},
\end{equation}
and
\begin{equation}
\boldsymbol{B}_1=
\begin{pmatrix}
\cos{(2\tau)} & 0 & 0 & 0 \\
\rho_1\cos{(2\varepsilon})\cos{(2\tau)} & 0 & 0 & -i\rho_2\sin{(2\varepsilon)} \\
\rho_2\cos{(2\varepsilon)}\cos{(2\tau)} & 0 & 0 & -i\rho_1\sin{(2\varepsilon)} \\
\rho_3\cos{(2\tau)} & 0 & 0 & 0
\end{pmatrix},\qquad 
\boldsymbol{B}_2=
\begin{pmatrix}
0 & 1 & 0 & 0 \\
0 & \rho_1\cos{(2\varepsilon)} & 0 & 0 \\
0 & \rho_2\cos{(2\varepsilon)}& 0 & 0 \\
0 & \rho_3 & 0 & 0
\end{pmatrix}.
\end{equation}
By again following the same steps outlined in the proof of Proposition \ref{lem:1}, we obtain
\begin{equation}
\label{48}
D'(n)=\sum_{m=0}^3\left[(\boldsymbol{B}_1-i\sin{(2\tau)}\boldsymbol{B}_2)^n\right]_{0m}\rho_m,
\end{equation}
which depends only on the matrices $\boldsymbol{B}_{1,2}$. To further simplify this expression one can set $\rho_1=1$, $\rho_2=\rho_3=0$, corresponding to when each ancilla is initalized in the state $\rho_A = |+\rangle\langle +|_A$, \eqref{rhoa}. We then get ($D'(0)=1$)
\begin{equation}
D'(n)= \e^{-2i\tau} (\cos{(2\tau)}-i\sin{(2\tau)}\cos{(2\varepsilon)})^{n-1}, \quad n\ge 1,
\end{equation}
giving \eqref{d'n}.

\bigskip

{\bf Acknowledgements.} This work was supported by an Alliance International Catalyst Grant offered by the Natural Sciences and Engineering Research Council of Canada (NSERC). The work of A.N. and M.M. was additionally supported by an NSERC Discovery Grant. M.M. thanks the University of Stellenbosch for the hospitality received during a visit when this work was initiated, and G.P. is grateful to Memorial University for hosting him during the research phase. The authors thank two anonymous referees for examining this work and suggesting several improvements.


\begin{thebibliography}{99}

\bibitem{Latune2023}
C.~L. Latune, G.~Pleasance, and F.~Petruccione, {\it Cyclic Quantum Engines Enhanced by Strong Bath Coupling}, Phys.~Rev.~Applied~{\bf 20}, 024038 (2023). 

\bibitem{Pleasance2024}
G.~Pleasance and F.~Petruccione, {\it Nonequilibrium Quantum Heat Transport between Structured Environments}, New~J.~Phys. {\bf 26}, 73025 (2024).

\bibitem{Breuer2002}
H.-P.~Breuer and F.~Petruccione, {\it The Theory of Open Quantum Systems} (Oxford University Press, New York, 2002).

\bibitem{Gorini1976}
V.~Gorini, A.~Kossakowski, and E.~C.~G.~Sudarshan, {\it Completely Positive Dynamical Semigroups of N-Level Systems}, J.~Math.~Phys.~{\bf 17}, 821 (1976).

\bibitem{Lindblad1976}
G.~Lindblad, {\it On the Generators of Quantum Dynamical Semigroups}, Comm.~Math.~Phys. {\bf 48}, 119 (1976).

\bibitem{Johansson2013}
J.~R. Johansson, P.~D. Nation, and F.~Nori, {\it QuTiP 2: A Python Framework for the Dynamics of Open Quantum Systems}, Comput.~Phys.~Commun. {\bf 184}, 1234 (2013).

\bibitem{Plenio1998}
M.~B. Plenio and P.~L. Knight, {\it The Quantum-Jump Approach to Dissipative Dynamics in Quantum Optics}, Rev.~Mod.~Phys. {\bf 70}, 101 (1998).

\bibitem{Dalibard1992}
J.~Dalibard, Y.~Castin, and K.~M{\o}lmer, {\it Wave-Function Approach to Dissipative Processes in Quantum Optics}, Phys.~Rev.~Lett. {\bf 68}, 580 (1992).

\bibitem{Gisin1992}
N.~Gisin and I.~C. Percival, {\it The Quantum-State Diffusion Model Applied to Open Systems}, J.~Phys.~A: Math.~Gen. {\bf 25}, 5677 (1992).

\bibitem{Vega2017}
I.~de~Vega and D.~Alonso, {\it Dynamics of Non-Markovian Open Quantum Systems}, Rev.~Mod.~Phys. {\bf 89}, 15001 (2017).

\bibitem{Tanimura1990}
Y.~Tanimura, {\it Nonperturbative Expansion Method for a Quantum System Coupled to a Harmonic-Oscillator Bath}, Phys.~Rev.~A~{\bf 41}, 6676 (1990).

\bibitem{Ishizaki2005}
A.~Ishizaki and Y.~Tanimura, {\it Quantum Dynamics of System Strongly Coupled to Low-Temperature Colored Noise Bath: Reduced Hierarchy Equations Approach}, J.~Phys.~Soc.~Japan {\bf 74}, 3131 (2005).

\bibitem{Tamascelli2018}
D.~Tamascelli, A.~Smirne, S.~F~Huelga, and M.~B.~Plenio, {\it Nonperturbative Treatment of Non-Markovian Dynamics of Open Quantum Systems}, Phys.~Rev.~Lett.~{\bf 120}, 30402 (2018).

\bibitem{Pleasance2017}
G.~Pleasance and B.~M.~Garraway, {\it Application of Quantum Darwinism to a Structured Environment}, Phys.~Rev.~A~{\bf 96}, 62105 (2017).

\bibitem{Pleasance2020}
G.~Pleasance, B.~M.~Garraway, and F.~Petruccione, {\it Generalized Theory of Pseudomodes for Exact Descriptions of Non-Markovian Quantum Processes}, Phys.~Rev.~Res.~{\bf 2}, 43058 (2020).

\bibitem{Pleasance2021}
G.~Pleasance and F.~Petruccione, {\it Pseudomode Description of General Open Quantum System Dynamics: Non-Perturbative Master Equation for the Spin-Boson Model}, arXiv:2108.05755.

\bibitem{Albarelli2024}
F.~Albarelli, B.~Vacchini, and A.~Smirne, {\it Pseudomode Treatment of Strong-Coupling Quantum Thermodynamics}, Quantum~Sci.~Technol. {\bf 10}, 15041 (2024).

\bibitem{Weiss2011}
U. Weiss, {\it Quantum Dissipative Systems} (World Scientific, 2011).

\bibitem{Stenius1996}
P.~Stenius and A.~Imamoglu, {\it Stochastic Wavefunction Methods beyond the Born - Markov and Rotating-Wave Approximations}, Quantum~Semiclass.~Opt.~{\bf 8}, 283 (1996).

\bibitem{Piilo2008}
J.~Piilo, S.~Maniscalco, K.~H\"{a}rk\"{o}nen, and K.-A.~Suominen, {\it Non-Markovian Quantum Jumps}, Phys. Rev. Lett.~{\bf 100}, 180402 (2008).

\bibitem{Suess2014}
D.~Suess, A.~Eisfeld, and W.~T.~Strunz, {\it Hierarchy of Stochastic Pure States for Open Quantum System Dynamics}, Phys.~Rev.~Lett.~{\bf 113}, 150403 (2014).

\bibitem{Rau1963}
J.~Rau, {\it Relaxation Phenomena in Spin and Harmonic Oscillator Systems}, Phys.~Rev.~{\bf 129}, 1880 (1963).

\bibitem{Ciccarello2022}
F.~Ciccarello, S.~Lorenzo, V.~Giovannetti, and G.~Massimo Palma, {\it Quantum Collision Models: Open System Dynamics from Repeated Interactions}, Phys.~Rep.~{\bf 954}, 1 (2022).

\bibitem{Rybar2012}
T.~Ryb\'{a}r, S.~N.~Filippov, M.~Ziman, and V.~Bu\v{z}ek, {\it Simulation of Indivisible Qubit Channels in Collision Models}, J.~Phys.~B: At.~Mol.~Opt.~Phys. {\bf 45}, 154006 (2012).

\bibitem{Filippov2017}
S.~N.~Filippov, J.~Piilo, S.~Maniscalco, and M.~Ziman, {\it Divisibility of Quantum Dynamical Maps and Collision Models}, Phys.~Rev.~A {\bf 96}, 32111 (2017).

\bibitem{Cattaneo2021}
M.~Cattaneo, G.~De~Chiara, S.~Maniscalco, R.~Zambrini, and G. L.~Giorgi, {\it Collision Models Can Efficiently Simulate Any Multipartite Markovian Quantum Dynamics}, Phys.~Rev.~Lett. {\bf 126}, 130403 (2021).

\bibitem{Ciccarello2017}
F.~Ciccarello, {\it Collision Models in Quantum Optics}, Quantum Measurements and Quantum Metrology {\bf 4}, (2017).

\bibitem{Whalen2019}
S.~J.~Whalen, {\it Collision Model for Non-Markovian Quantum Trajectories}, Phys.~Rev.~A {\bf 100}, 52113 (2019).

\bibitem{Cuevas2019}
\'{A}.~Cuevas, A.~Geraldi, C.~Liorni, L.~D.~Bonavena, A.~De~Pasquale, F.~Sciarrino, V.~Giovannetti, and P.~Mataloni, {\it All-Optical Implementation of Collision-Based Evolutions of Open Quantum Systems}, Sci.~Rep.~{\bf 9}, (2019).

\bibitem{Campbell2019}
S.~Campbell, B.~\c{C}akmak, \"{O}. E. M\"{u}stecapl{\i}o{\u{g}}lu, M.~Paternostro, and B.~Vacchini, {\it Collisional Unfolding of Quantum Darwinism}, Phys.~Rev.~A~{\bf 99}, 42103 (2019).

\bibitem{Lorenzo2020}
S.~Lorenzo, M.~Paternostro, and G.~M.~Palma, {\it Reading a Qubit Quantum State with a Quantum Meter: Time Unfolding of Quantum Darwinism and Quantum Information Flux}, 	arXiv:2001.11558.

\bibitem{Lorenzo2015}
S.~Lorenzo, R.~McCloskey, F.~Ciccarello, M.~Paternostro, and G.~M.~Palma, {\it Landauer’s Principle in Multipartite Open Quantum System Dynamics}, Phys.~Rev.~Lett.~{\bf 115}, 120403 (2015).

\bibitem{Strasberg2017}
P.~Strasberg, G.~Schaller, T.~Brandes, and M.~Esposito, {\it Quantum and Information Thermodynamics: A Unifying Framework Based on Repeated Interactions}, Phys.~Rev.~X~{\bf 7}, 21003 (2017).

\bibitem{Morrone2023}
D.~Morrone, M.~A.~C. Rossi, A.~Smirne, and M.~G.~Genoni, {\it Charging a Quantum Battery in a Non-Markovian Environment: A Collisional Model Approach}, Quantum~Sci.~Technol.~{\bf 8}, 35007 (2023).

\bibitem{Cusumano2023}
S.~Cusumano and G.~D.~Chiara, {\it Structured Quantum Collision Models: Generating Coherence with Thermal Resources}, New~J.~Phys.~{\bf 26}, 23001 (2023).

\bibitem{Ribeiro2024}
W.~L.~Ribeiro and C.~A.~Moura, {\it Collisional Approach for Open Neutrino Systems}, arXiv:2411.08330.

\bibitem{Pellegrini2009}
C.~Pellegrini and F.~Petruccione, {\it Non-Markovian Quantum Repeated Interactions and Measurements}, J.~Phys.~Math.~Theor.~{\bf 42}, 425304 (2009).

\bibitem{Ciccarello2013}
F.~Ciccarello, G.~M.~Palma, and V.~Giovannetti, {\it Collision-Model-Based Approach to Non-Markovian Quantum Dynamics}, Phys.~Rev.~A~{\bf 87}, 40103 (2013).

\bibitem{Ciccarello2013a}
F.~Ciccarello and V.~Giovannetti, {\it A Quantum Non-Markovian Collision Model: Incoherent Swap Case}, Phys.~Scripta~{\bf T153}, 14010 (2013).

\bibitem{McCloskey2014}
R.~McCloskey and M.~Paternostro, {\it Non-Markovianity and System-Environment Correlations in a Microscopic Collision Model}, Phys.~Rev.~A~{\bf 89}, 52120 (2014).

\bibitem{Kretschmer2016}
S.~Kretschmer, K.~Luoma, and W.~T.~Strunz, {\it Collision Model for Non-Markovian Quantum Dynamics}, Phys.~Rev.~A~{\bf 94}, 12106 (2016).

\bibitem{Bernardes2014}
N.~K.~Bernardes, A.~R.~R.~Carvalho, C.~H.~Monken, and M.~F.~Santos, {\it Environmental Correlations and Markovian to Non-Markovian Transitions in Collisional Models}, Phys.~Rev.~A~{\bf 90}, 32111 (2014).

\bibitem{Bernardes2017}
N.~K.~Bernardes, A.~R.~R.~Carvalho, C.~H.~Monken, and M.~F.~Santos, {\it Coarse Graining a Non-Markovian Collisional Model}, Phys.~Rev.~A~{\bf 95}, 32117 (2017).

\bibitem{Mascarenhas2017}
E.~Mascarenhas and I.~de~Vega, {\it Quantum Critical Probing and Simulation of Colored Quantum Noise}, Phys.~Rev.~A~{\bf 96}, 62117 (2017).

\bibitem{Lorenzo2017}
S.~Lorenzo, F.~Ciccarello, and G.~M.~Palma, {\it Composite Quantum Collision Models}, Phys.~Rev.~A~{\bf 96}, 32107 (2017).

\bibitem{Campbell2018}
S.~Campbell, F.~Ciccarello, G.~M.~Palma, and B.~Vacchini, {\it System-Environment Correlations and Markovian Embedding of Quantum Non-Markovian Dynamics}, Phys.~Rev.~A~{\bf 98}, 12142 (2018).

\bibitem{Bernardes2016}
N.~K.~Bernardes, J.~P.~S.~Peterson, R.~S.~Sarthour, A.~M.~Souza, C.~H.~Monken, I.~Roditi, I.~S.~Oliveira, and M.~F.~Santos, {\it High Resolution Non-Markovianity in NMR}, Sci.~Rep.~{\bf 6}, 33945 (2016).

\bibitem{Jin2015}
J. Jin, V. Giovannetti, R. Fazio, F. Sciarrino, P. Mataloni, A. Crespi, and R. Osellame, {\it All-Optical Non-Markovian Stroboscopic Quantum Simulator}, Phys.~Rev.~A~{\bf 91}, 12122 (2015).

\bibitem{Garcia2020}
G. Garc\'{i}-P\'{e}rez, M. A. C. Rossi, and S. Maniscalco, {\it IBM Q Experience as a Versatile Experimental Testbed for Simulating Open Quantum Systems}, npj~Quantum~Inf.~{\bf 6}, 1 (2020).

\bibitem{Man2018}
Z.-X.~Man, Y.-J.~Xia, and R.~Lo~Franco, {\it Temperature Effects on Quantum Non-Markovianity via Collision Models}, Phys.~Rev.~A~{\bf 97}, 62104 (2018).

\bibitem{Whalen2017}
S.~J.~Whalen, A.~L.~Grimsmo, and H.~J.~Carmichael, {\it Open Quantum Systems with Delayed Coherent Feedback}, Quantum~Sci.~Technol.~{\bf 2}, 44008 (2017).

\bibitem{Li2021}
H.~Li, J.~Zou, and B.~Shao, {\it Enhanced Quantumness via Non-Markovianity}, Phys.~Rev.~A~{\bf 104}, 52201 (2021).

\bibitem{Saha2024}
T.~Saha, A.~Das, and S.~Ghosh, {\it Quantum Homogenization in Non-Markovian Collisional Model}, New~J.~Phys.~{\bf 26}, 23011 (2024).

\bibitem{Saha2024a}
T. Saha, Sahil, K. P. Athulya, and S. Ghosh, {\it Post-Markovian Master Equation \`{A} La Microscopic Collisional Model}, arXiv:2411.16878.

\bibitem{Wolf2008}
M.~M.~Wolf and J.~I.~Cirac, {\it Dividing Quantum Channels}, Commun.~Math.~Phys.~{\bf 279}, 147 (2008).

\bibitem{Rivas2014}
\'{A}.~Rivas, S.~F.~Huelga, and M.~B.~Plenio, {\it Quantum Non-Markovianity: Characterization, Quantification and Detection}, Rep.~Prog.~Phys.~{\bf 77}, 94001 (2014).

\bibitem{Breuer2009}
H.-P.~Breuer, E.-M.~Laine, and J.~Piilo, {\it Measure for the Degree of Non-Markovian Behavior of Quantum Processes in Open Systems}, Phys.~Rev.~Lett.~{\bf 103}, 210401 (2009).

\bibitem{Laine2010}
E.-M.~Laine, J.~Piilo, and H.-P.~Breuer, {\it Measure for the Non-Markovianity of Quantum Processes}, Phys.~Rev.~A~{\bf 81}, 62115 (2010).

\bibitem{Wissmann2012}
S.~Wi\ss mann, A.~Karlsson, E.-M.~Laine, J.~Piilo, and H.-P.~Breuer, {\it Optimal State Pairs for Non-Markovian Quantum Dynamics}, Phys.~Rev.~A~{\bf 86}, 62108 (2012).

\bibitem{Wootters1998}
W.~K.~Wootters, {\it Entanglement of Formation of an Arbitrary State of Two Qubits}, Phys.~Rev.~Lett.~{\bf 80}, 2245 (1998).

\bibitem{Nielsen2012}
M.~A.~Nielsen and I.~L.~Chuang, {\it Quantum Computation and Quantum Information: 10th Anniversary Edition} (Cambridge University Press, 2012).

\bibitem{Coffman2000} V.~Coffman, J.~Kundu, and W.~K.~Wootters, {\it Distributed Entanglement}, Phys.~Rev.~A~{\bf 61}, 052306 (2000).

\bibitem{Osborne2006} T.~J. Osborne, F.~Verstraete, {\it General Monogamy Inequality for Bipartite Qubit Entanglement}, Phys.~Rev.~Lett.~{\bf 96}, 220503 (2006).

\bibitem{Laine2010b} E.-M. Laine, J.~Piilo, and H.-P.~Breuer, {\it Witness for Initial System-Environment Correlations in Open-System Dynamics}, Europhys.~Lett.~{\bf 92}, 60010 (2010).

\bibitem{Breuer2016} H.-P.~Breuer, E.-M.~Laine, J.~Piilo, and B.~Vacchini, {\it Colloquium: Non-Markovian Dynamics in Open Quantum Systems}, Rev.~Mod.~Phys.~{\bf 88}, 21002 (2016).

\bibitem{PleasanceTBC} G.~Pleasance, \'{A}.~E.~Neira, M.~Merkli, and F.~Petruccione, in preparation.  

\bibitem{DeChiara2020}
G.~De~Chiara and M.~Antezza, {\it Quantum Machines Powered by Correlated Baths}, Phys.~Rev.~Res.~{\bf 2}, 33315 (2020).

\bibitem{Stinespring1955}
W.~F.~Stinespring, {\it Positive Functions on $C^*$-Algebras}, Proc.~Amer.~Math.~Soc.~{\bf 6}, 211 (1955).

\end{thebibliography}
\end{document}